\documentclass[aps,pre,floatfix,superscriptaddress,preprint,10pt]{revtex4}

\usepackage{bm}
\usepackage{empheq}
\usepackage{float}
\usepackage{xcolor}
\usepackage{amsfonts}
\usepackage{amssymb}
\usepackage{amsmath}
\usepackage[capitalise,noabbrev]{cleveref}
\usepackage{appendix}
\usepackage{multirow}
\usepackage{physics}
\usepackage{dutchcal}
\usepackage{graphicx}

\allowdisplaybreaks
\everymath={\displaystyle}
\crefname{equation}{}{}
\crefrangelabelformat{equation}{(#3#1#4)--(#5\crefstripprefix{#1}{#2}#6)}

\newcommand\numberthis{\refstepcounter{equation}\tag{\theequation}}
\newcommand{\etal}{{\it et al.\ }}
\newcommand{\ie}{i.e.\ }
\newcommand{\eg}{e.g.\ }
\newcommand{\Oh}{O}

\newcommand{\n}{\bm{n}}
\newcommand{\vel}{\bm{u}}
\newcommand{\tp}{\tilde{p}}

\newcommand{\unitx}{\bm{\hat{x}}}
\newcommand{\unity}{\bm{\hat{y}}}
\newcommand{\unitz}{\bm{\hat{z}}}
\newcommand{\thetaL}{\theta_{\rm L}}

\newcommand{\wf}{W_{\rm F}}
\newcommand{\Disp}{\mathcal{D}}

\newcommand{\Thetat}{\Theta_{\rm c}}
\newcommand{\Phic}{\Phi_{\rm c}}
\newcommand{\Phir}{\Phi_{\rm c}}

\newcommand{\xr}{\hat{x}}
\newcommand{\yr}{\hat{y}}
\newcommand{\hPsi}{\hat{\Psi}}
\newcommand{\gc}{g_{\rm c}}
\newcommand{\hc}{h_{\rm c}}
\newcommand{\mg}{\mathcal{g}}
\newcommand{\mh}{\mathcal{h}}
\newcommand{\gx}{g_1}
\newcommand{\gy}{{g}_2}
\newcommand{\gxy}{{g}_3}
\newcommand{\mgx}{\mathcal{g}_1}
\newcommand{\mgy}{\mathcal{g}_2}
\newcommand{\mgxy}{\mathcal{g}_3}
\newcommand{\bgx}{\mathcal{b}_{1}}
\newcommand{\bgy}{\mathcal{b}_{2}}
\newcommand{\bgxy}{\mathcal{b}_{3}}
\newcommand{\cg}{g_{\rm c}}
\newcommand{\ch}{h_{\rm c}}
\newcommand{\cgx}{g_{\rm 1c}}
\newcommand{\cgy}{g_{\rm 2c}}
\newcommand{\cgxy}{g_{\rm 3c}}
\newcommand{\ER}{\text{Er}}
\newcommand{\RE}{\text{Re}}

\newcommand{\etaL}{\eta_{\rm L}}
\newcommand{\Sp}{s_{\rm p}}

\newcommand{\Hscale}{D}
\newcommand{\Hvscale}{S}

\begin{document}

\title{Hele-Shaw flow of a nematic liquid crystal}
\author{Joseph R.\ L.\ Cousins} 
\email{joseph.cousins@strath.ac.uk}
\affiliation{Department of Mathematics and Statistics, University of Strathclyde, Livingstone Tower, 26 Richmond Street, Glasgow G1 1XH, United Kingdom}
\affiliation{School of Mathematics and Statistics, University of Glasgow, University Place, Glasgow G12 8QQ, United Kingdom}
\author{Nigel J.\ Mottram} 
\email{nigel.mottram@glasgow.ac.uk}
\affiliation{School of Mathematics and Statistics, University of Glasgow, University Place, Glasgow G12 8QQ, United Kingdom}
\author{Stephen K.\ Wilson} 
\thanks{Author for correspondence}
\email{s.k.wilson@strath.ac.uk}
\affiliation{Department of Mathematics and Statistics, University of Strathclyde, Livingstone Tower, 26 Richmond Street, Glasgow G1 1XH, United Kingdom}

\date{23rd January 2024}

\begin{abstract}
Motivated by the variety of applications in which nematic Hele-Shaw flow occurs, a theoretical model for Hele-Shaw flow of a nematic liquid crystal is formulated and analysed.
We derive the thin-film Ericksen–Leslie equations that govern nematic Hele-Shaw flow, and consider two important limiting cases in which we can make significant analytical progress.
Firstly, we consider the leading-order problem in the limiting case in which elasticity effects dominate viscous effects, and find that the nematic liquid crystal anchoring on the plates leads to a fixed director field and an anisotropic patterned viscosity that can be used to guide the flow of the nematic.
Secondly, we consider the leading-order problem in the opposite limiting case in which viscous effects dominate elasticity effects, and find that the flow is identical to that of an isotropic fluid and the behaviour of the director is determined by the flow.
As an example of the insight which can be gained by using the present approach, we then consider the flow of nematic according to a simple model for the squeezing stage of the One Drop Filling method, an important method for the manufacture of Liquid Crystal Displays, in these two limiting cases.
\end{abstract}

\maketitle

\section{Introduction}
\label{sec:intro}

Following the original experiments by Hele-Shaw \cite{HeleShaw1898} and the pioneering theory for the flow of a viscous fluid by Stokes \cite{Stokes1898}, interest in what is now termed Hele-Shaw flow has remained the subject of ongoing research for well over a century.
What is now called a Hele-Shaw cell consists of two parallel plates separated by a narrow gap which is partially or wholly filled with viscous fluid.
Mathematically, this system naturally lends itself to a thin-film (\ie a lubrication) analysis, and significant progress is often possible using analytical methods and reduced models that are computationally much cheaper than fully numerical alternatives \cite{AchesonBOOK}.
Experimentally, Hele-Shaw cells are a useful tool for visualising two-dimensional flows that have allowed researchers to investigate many fluid mechanical effects.
Such effects include viscous fingering \cite{SaffmanTaylor1958,Saffman1986}, porosity \cite{Howison:1986,Huppert/Woods:1995,Homsy:1987}, and bubble dynamics \cite{Kopfsill/Homsy:1988b,Tanveer/Saffman:1987}.
An extensive list of work up to 1998 that details many of the applications of isotropic Hele-Shaw cells is available at \cite{HowisonURL}, and a more up-to-date review of Hele-Shaw flow is given by Morrow \etal \cite{Morrow2021}.

Although much of the work on Hele-Shaw flow has focused on isotropic fluids, there has also been some interest in the Hele-Shaw flow of non-Newtonian fluids, including the flow of power-law fluids by Hassager and Lauridsen \cite{Hassager/Lauridsen:1988} and the flow of viscoelastic fluids by Ro and Homsy \cite{Ro/Homsy:1995}.
Somewhat surprisingly, there has been little work on the theory of Hele-Shaw cells filled with liquid crystals, despite their relevance to the liquid crystal display (LCD) industry, in which thin-film flows of liquid crystal between parallel plates are a key element of device manufacture \cite{CapillaryMiYang}.
Liquid crystals are anisotropic fluids with long-range molecular orientational order and possibly molecular positional order that exhibit a rich variety of physical behaviours, including anisotropic elasticity, viscosity and surface effects \cite{ISBOOK}.
The most common type of liquid crystal used in LCDs is a nematic liquid crystal (nematic), which exhibit orientational order but no positional order.
The standard continuum approach used to mathematically model the behaviour of nematics makes use of the so-called director $\n$, a unit vector representing the average nematic molecular orientation, together with the fluid velocity $\vel$ and fluid pressure $p$, to formulate the conservation of mass equation, the conservation of linear momentum equations, and the conservation of angular momentum equations.
These conservation equations are known as the Ericksen--Leslie equations \cite{EL1,EL2}, and they have been successfully applied to a variety of problems involving the flow of nematics \cite{ISBOOK,GPBOOK}.

Although there has been only limited theoretical study of nematic Hele-Shaw flow, there has been some experimental work on nematic viscous fingering \cite{Buka1986,Sonin1993, Lam1989,Lam1991}, some of which has included simple theoretical models \cite{Lam1989,Lam1991}.
For example, Lam \etal \cite{Lam1989} consider a number of fixed director fields for which the flow of a nematic is identical to the flow of an isotropic fluid with effective viscosity determined by the fixed director field.
Following the initial experiments on nematic viscous fingering by Buka \etal \cite{Buka1986} and Sonin and Bartolino \cite{Sonin1993}, there have been a variety of extensions of this work, for example, to the viscous fingering of nematics under applied electric fields by Folch \etal \cite{Folch2000,Folch2001} and T\'oth-Katona and Buka \cite{TothKatona2003}.
Nematic microfluidic experiments have also been the topic of much recent interest \cite{Sengupta2014}, in particular, experiments by Sengupta \etal \cite{Sengupta2011,Sengupta2012,Sengupta2013} have inspired work on a variety of effects, including control of nematic defects \cite{Giomi2017,Sengupta2013b,Bhadwal2020}, micropillar induced cavitation \cite{Sengupta2013d,Stieger2017}, and control of nematic flow using external stimuli \cite{Sengupta2013c,Na2010}.
These experiments have also initiated a number of theoretical investigations.
For example, flow transitions observed by Sengupta \etal \cite{Sengupta2011,Sengupta2012,Sengupta2013} have been studied using the Ericksen--Leslie equations by Anderson \etal \cite{Anderson2015} and Crespo \etal \cite{Crespo2017} and using lattice Boltzmann simulations by Batista \etal \cite{Batista2015}.
However, many of these studies employ one-dimensional models for the nematic that cannot capture two-dimensional effects.

As previously mentioned, Hele-Shaw flow is particularly relevant to the industrial manufacture of LCDs, which involves filling the gap between parallel plates with nematic \cite{ODFCapillary}.
LCD manufacturing is currently carried out using one of two methods: the capillary-filling method and the One Drop Filling (ODF) method \cite{HANDBOOK_ODF}.
In the capillary-filling method, the nematic is introduced at one edge of the parallel plates and fills the gap between the plates via capillary action, which results in relatively low flow speeds and long manufacturing times \cite{CapillaryMiYang}.
In the ODF method, an array of nematic drops are dispensed on one plate, and a second (parallel) plate is lowered onto droplets, squeezing them until they coalesce to form a continuous nematic film between the plates, which results in relatively high flow speeds and short manufacturing times \cite{HANDBOOK_ODF}.
The ODF method is often preferred because of this higher manufacturing throughput of devices when compared to the capillary-filling method.
However, the ODF method can sometimes lead to unwanted optical effects, known as ODF mura, which degrade the quality of the final display \cite{MuraLee,MuraPratt,Cousins2018}.
Somewhat surprisingly, there has been relatively little work to model the nematic flow in these manufacturing methods using the standard theoretical approach for isotropic Hele-Shaw flow \cite{AchesonBOOK}.
For further discussion of nematic flow in LCD manufacturing see Cousins \etal \cite{Cousinsthesis,Cousins2019,Cousins2020}.

Motivated by the variety of applications in which nematic Hele-Shaw flow occurs, in the present work we formulate and analyse a theoretical model for Hele-Shaw flow of a nematic liquid crystal.
In \cref{sec:prob,sec:EL,sec:BC,sec:BC:fs,sec:ND,sec:thinfilm}, we derive the thin-film Ericksen–Leslie equations that govern nematic Hele-Shaw flow.
Then, we consider these equations in a number of important limiting cases in which we can make significant analytical progress.
Firstly, in \cref{sec:ER0}, we consider the leading-order problem in the limiting case in which elasticity effects dominate viscous effects and, secondly, in \cref{sec:ERinf}, we consider the leading-order problem in the opposite limiting case in which viscous effects dominate elasticity effects.
Finally, in \cref{sec:app}, as an example of the insight which can be gained by using the present approach, we consider the flow of nematic according to a simple model for the squeezing stage of the ODF method in these two limiting cases.

\section{Problem formulation}
\label{sec:prob}

\begin{figure}[tp]
\begin{center}
\includegraphics[width=0.7\linewidth]{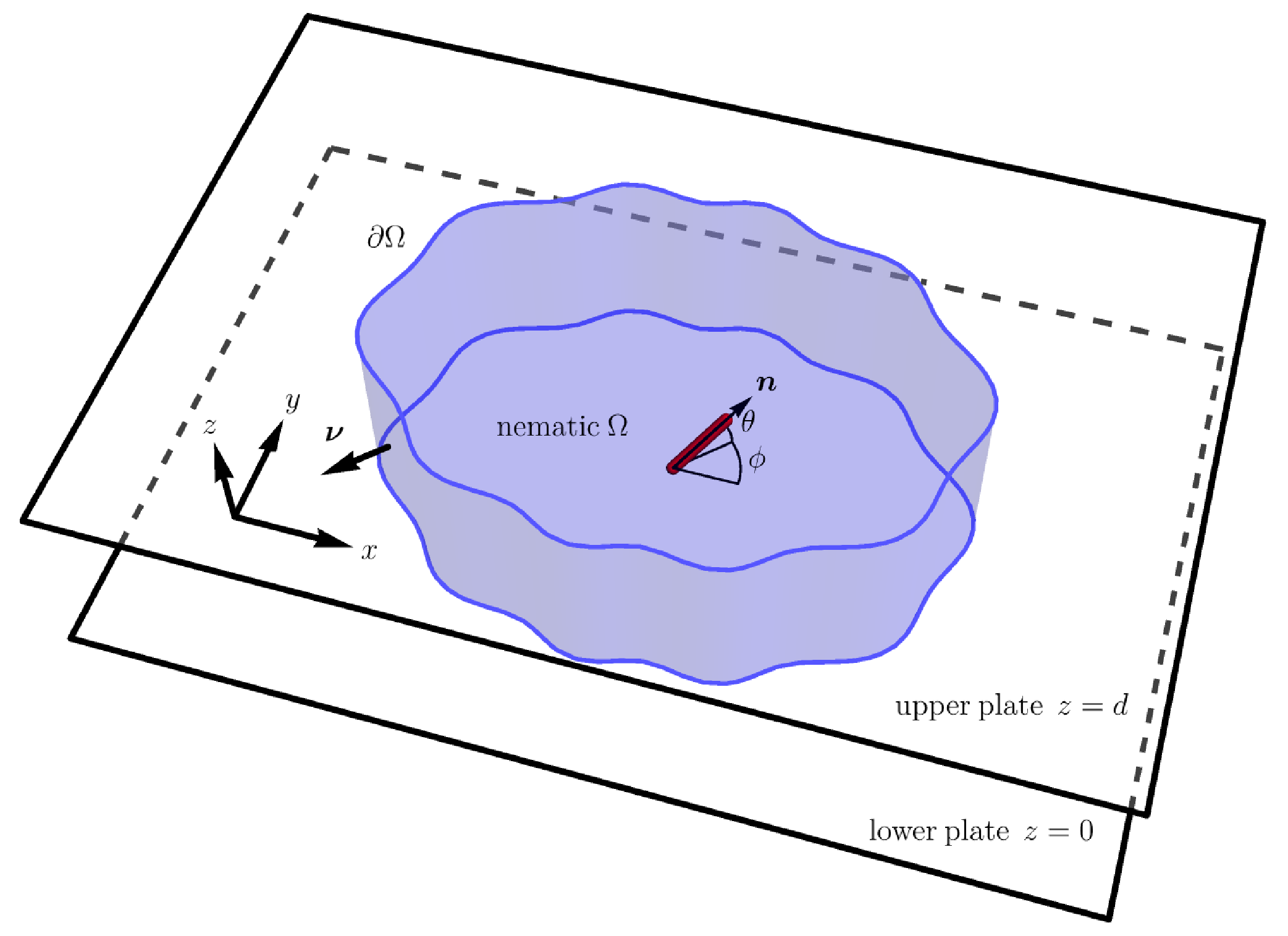}
\end{center}
\caption{
A Hele-Shaw cell showing a perspective view of a region of nematic $\Omega$ (in light blue) bounded between solid parallel plates at $z=0$ and $z=d$, and with a free surface $\partial \Omega$ with outward unit normal $\bm{\nu}$.
The Cartesian coordinates $x$, $y$ and $z$, the director $\n$, the tilt director angle $\theta$, and the twist director angle $\phi$ are also shown.
}
\label{Figure1}
\end{figure}

We consider the flow of a nematic in a standard Hele-Shaw cell that consists of two parallel plates separated by a narrow gap which is partially or wholly filled with the nematic.
In particular, we consider a region of nematic $\Omega=\Omega(t)$ bounded between solid parallel plates at $z=0$ (which we term the lower plate) and $z=d(t)$ (which we term the upper plate), and with a free surface $\partial \Omega=\partial \Omega(t)$ with outward unit normal $\bm{\nu} = \bm{\nu}(x,y,z,t)$, where $x$, $y$ and $z$ are Cartesian coordinates, $t$ denotes time, as shown in \cref{Figure1}. Note that the thickness of the cell in the $z$-direction, $d(t)$, may, in general, be time dependent, and so we allow for the possibility of the upper plate to move in the direction perpendicular to the plates, with the velocity of the upper plate denoted $d'=d'(t)$. (Note that, despite what \cref{Figure1} might suggest, in the following mathematical model it is not necessary for the region of nematic to be simply connected.)
The nematic velocity and the fluid pressure are denoted by
\begin{align}
    \vel &= u(x,y,z,t) \unitx + v(x,y,z,t) \unity + w(x,y,z,t)\unitz, \label{velocity} \\
    p &= p(x,y,z,t), \label{pressure}
\end{align}
respectively, where
$\unitx$, $\unity$ and $\unitz$ are the Cartesian coordinate unit vectors in the $x$-, $y$- and $z$-directions and $u$, $v$ and $w$ are the components of the velocity in the $x$-, $y$- and $z$-directions.
The director $\n=\n(x,y,z,t)$ is written in the form
\begin{align}
    \n &= \cos\theta(x,y,z,t)\cos\phi(x,y,z,t) \, \unitx + \cos\theta(x,y,z,t)\sin\phi(x,y,z,t) \, \unity +\sin\theta(x,y,z,t) \, \unitz, \label{director}
\end{align}
where $\theta(x,y,z,t)$ is the angle between the director and the $xy$-plane, which is commonly called the tilt director angle, and $\phi(x,y,z,t)$ is the angle between the projection of the director onto the $xy$-plane, and the $x$-axis, which is commonly called the twist director angle \cite{ISBOOK}.

\section{The governing equations}
\label{sec:EL}

In this work we consider governing equations for the velocity, pressure and director in the form of the Ericksen--Leslie equations \cite{EL1,EL2}, which are derived from a continuum approach and the principles of the conservation of mass, the conservation of linear momentum and the conservation of angular momentum.

The conservation of mass equation is given by
\begin{align}
   \pdv{u}{x}+ \pdv{v}{y} +\pdv{w}{z} = 0. \label{CM}
\end{align}

The conservation of linear momentum equations are given by
\begin{align}
   \rho \dot{u} &= - \pdv{\tp}{x} + \pdv{x}\left(\pdv{\Disp}{u_x}\right)+\pdv{y}\left(\pdv{\Disp}{u_y}\right)+\pdv{z}\left(\pdv{\Disp}{u_z}\right) - \pdv{\Disp}{\theta} \pdv{\theta}{x}- \pdv{\Disp}{\phi} \pdv{\phi}{x}, \label{LM1} \\
   \rho \dot{v} &= - \pdv{\tp}{y} + \pdv{x}\left(\pdv{\Disp}{v_x}\right)+\pdv{y}\left(\pdv{\Disp}{v_y}\right)+\pdv{z}\left(\pdv{\Disp}{v_z}\right) - \pdv{\Disp}{\theta} \pdv{\theta}{y}- \pdv{\Disp}{\phi} \pdv{\phi}{y}, \label{LM2} \\
   \rho \dot{w} &= - \pdv{\tp}{z} + \pdv{x}\left(\pdv{\Disp}{w_x}\right)+\pdv{y}\left(\pdv{\Disp}{w_y}\right)+\pdv{z}\left(\pdv{\Disp}{w_z}\right) - \pdv{\Disp}{\theta} \pdv{\theta}{z}- \pdv{\Disp}{\phi} \pdv{\phi}{z}, \label{LM3}
\end{align}
where $\Disp = \Disp(\n,\bm{N},\bm{A})$ is the nematic viscous dissipation, which depends on the director $\n$, the co-rotational time flux of the director $\bm{N}=\partial {\n}/\partial t + (\vel \cdot \grad) \n - (\grad \vel - (\grad \vel)^{\rm T}) \, \n/2$ with $\grad = \unitx \, \partial/\partial {x} + \unity \, \partial/\partial {y} + \unitz \, \partial/\partial {z}$, and the rate of the strain tensor $\bm{A} = (\grad \vel +(\grad \vel)^{\rm T})/2$ \cite{ISBOOK}.
Also appearing in \cref{LM1,LM2,LM3} is the constant fluid density $\rho$ and the material time derivatives of $u$, $v$, and $w$ denoted by $\dot{u}$, $\dot{v}$, and $\dot{w}$, respectively, where, for example, $\dot{u} = \partial u/ \partial t + (\vel \cdot \grad) u$.
Additionally, in \cref{LM1,LM2,LM3}, $\tp$ is the modified pressure \cite{ISBOOK}, henceforth simply called the pressure for brevity, which can be expressed in terms of the fluid pressure $p$ as
\begin{align} \label{npressure}
    \tp = p +\wf-\hPsi,
\end{align}
where $\hPsi = \hPsi(x,y,z,t)$ is the bulk energy density corresponding to a general conservative body force \cite[Section 4.3]{ISBOOK}, for example body forces due to applied electric and/or magnetic fields and gravity, and $\wf=\wf(\n,\grad \n)$ is the nematic bulk elastic energy density, which depends on the director and spatial derivatives of the director.
In what follows, we leave $\hPsi$ unspecified to keep the approach as general as possible for now, but in \cref{sec:thinfilm}, for simplicity, we will neglect any conservative body forces and hence set $\hPsi \equiv 0$.

Finally, the conservation of angular momentum equations are given by
\begin{align}
  \pdv{\Disp}{\dot{\theta}} &= \pdv{x}\left(\pdv{\wf}{\theta_x}\right)+\pdv{y}\left(\pdv{\wf}{\theta_y}\right)+\pdv{z}\left(\pdv{\wf}{\theta_z}\right)-\pdv{\wf}{\theta}+\pdv{\hPsi}{\theta}, \label{AM1} \\
  \pdv{\Disp}{\dot{\phi}} &= \pdv{x}\left(\pdv{\wf}{\phi_x}\right)+\pdv{y}\left(\pdv{\wf}{\phi_y}\right)+\pdv{z}\left(\pdv{\wf}{\phi_z}\right)-\pdv{\wf}{\phi}+\pdv{\hPsi}{\phi}, \label{AM2}
\end{align}
where $\dot{\theta}$ and $\dot{\phi}$ are the material time derivatives of $\theta$ and $\phi$.
Together, the Ericksen--Leslie equations \cref{LM1,LM2,LM3,CM,AM1,AM2} with the unknowns $u$, $v$, $w$, $\tp$, $\theta$, and $\phi$, describe the behaviour in the bulk of the nematic.

To complete the Ericksen--Leslie equations, we now specify the form of the nematic viscous dissipation $\Disp$ and the nematic bulk elastic energy density $\wf$ appearing in \cref{LM1,LM2,LM3,AM1,AM2}.
In particular, we use of the standard nematic viscous dissipation $\Disp$ \cite{ISBOOK}, namely
\begin{align}
\Disp &= \dfrac{1}{2} \Big[\alpha_1 \left(\n \cdot \bm{A} \, \n \right)^2 +2(\alpha_6-\alpha_5) \bm{N} \cdot \bm{A} \, \n +\alpha_4 \: \text{tr}(\bm{A}^2) + \left(\alpha_5 +\alpha_6 \right) \big( \bm{A}\, \n \big)^2 + (\alpha_3-\alpha_2) \bm{N}^2 \Big]. \label{Disp}
\end{align}
The coefficients $\alpha_1,\ldots,\alpha_6$ appearing in \cref{Disp} are the Leslie viscosities \cite{ISBOOK}.
The expanded expression for the nematic viscous dissipation $\Disp$, using the director in the form of \cref{director}, the co-rotational time flux of the director, $\bm{N}$, and the rate of strain tensor, $\bm{A}$, is given in \cref{sec:appendixa}.
For later use, we note that the Leslie viscosities can be expressed in terms of the set of more easily measured nematic viscosities \cite{ISBOOK}, namely the rotational viscosity $\gamma_1$, the torsional viscosity $\gamma_2$, and the Miesowicz viscosities $\eta_1$, $\eta_2$, $\eta_3$ and $\eta_{12}$, as
\begin{align}\label{etagamma}
\begin{gathered}
    \gamma_1 = \alpha_3 - \alpha_2, \quad \gamma_2 = \alpha_6-\alpha_5 = \alpha_2 +\alpha_3, \\
    {\eta}_{1} = \dfrac{1}{2} \left( \alpha_3+\alpha_4+\alpha_6 \right), \quad
    {\eta}_{2} = \dfrac{1}{2} \left(-\alpha_2+\alpha_4+\alpha_5 \right), \quad
    {\eta}_{3} = \dfrac{1}{2} {\alpha}_4, \quad
    {\eta}_{12} = {\alpha}_1.
\end{gathered}
\end{align}
We take $\wf$ to be the Oseen--Frank bulk elastic energy density, which is defined by
\begin{align}
    \wf &= \dfrac{1}{2} K_1 (\grad \cdot \n)^2+\dfrac{1}{2} K_2 (\n \cdot \grad \times \n)^2+ \dfrac{1}{2} K_3 (\n \times \grad \times \n)^2 +\dfrac{1}{2} (K_2+K_4) \grad \cdot \big[(\n \cdot \grad) \n - ( \grad \cdot \n)\n\big], \label{WF}
\end{align}
where the constants $K_1$, $K_2$ and $K_3$ are the nematic splay, twist and bend elastic constants, respectively, and the combination $K_2+K_4$ is the saddle-splay elastic constant \cite{ISBOOK}.
To produce a mathematically tractable set of equations, we use the one-constant approximation, and so set $K_1=K_2=K_3=K$ and $K_4=0$, where $K$ is the one-constant elastic constant.
It is, in principle, possible to proceed without making the one-constant approximation, although the subsequent expressions become considerably more algebraically complicated.
In practice, the values of the elastic constants rarely differ by more than a factor of two, and so the one-constant approximation qualitatively describes the behaviour \cite{ISBOOK}.
Combining the one-constant approximation, the director in the form of \cref{director}, and $\wf$ defined by \cref{WF}, yields the one-constant approximation of the Oseen--Frank bulk elastic energy $\wf$ \cite{ISBOOK}, given by
\begin{align}
     \wf &= \dfrac{K}{2} \bigg[ \big(\grad\theta\big) ^2 + \cos^2\theta \,\big( \grad\phi\big)^2 \bigg]. \label{WF0}
\end{align}

\section{Boundary Conditions: nematic--plate interfaces}
\label{sec:BC}

At the interfaces between the nematic and the lower and upper plates, the boundary conditions for $\vel$ are standard no-slip and no-penetration conditions, namely
\begin{equation}
    u = v = w = 0 \quad \text{on} \quad z = 0 \label{noslip0}
\end{equation}
and
\begin{equation}
    u = v = 0 \quad \text{and} \quad w = d' \quad \text{on} \quad z = d. \label{noslipH}
\end{equation}

The boundary conditions on the director at the lower and upper plates are a result of intermolecular forces between the nematic and the material of which the plates are made. These forces can be produced through mechanical and/or chemical treatment of the plates to achieve a variety of desired preferred orientations \cite{SONINBOOK}. For example, photo-curable polymers embedded in the nematic or mechanical rubbing of the plates \cite{Chigrinov2021} have been used to create either a homogeneous preferred orientation or a patterned anchoring for which the preferred orientation varies in space. The resulting intermolecular forces lead to an energetically preferred orientation of the director at the nematic--plate interface, which are incorporated into boundary conditions for $\n$, called anchoring conditions.

In situations where the intermolecular forces are strong enough to prescribe the orientation of the director, the anchoring conditions are called infinite anchoring conditions \cite{SONINBOOK}.
When the intermolecular forces are weaker than this, the orientation of the director at the plates is also influenced by other effects in the system, such as the torque due to elasticity effects from the bulk of the nematic region.
This type of anchoring is known as weak anchoring, and the reorientation of the director away from the preferred orientation is known as anchoring breaking \cite{Cousins2023}.

One particular example of a preferred orientation of the director occurs when the intermolecular forces between the nematic and the plates are such that there is a preferred angle between the director and the plate normal, but the orientation of the director around the plate normal is not fixed. In this scenario the energetic preference is for the director to lie on a cone, a situation known as conical anchoring. Conical anchoring may be infinite, when intermolecular forces are strong enough to prescribe the angle of the director relative to the plate normal, or weak, when the orientation of the director at the plates is also influenced by other effects in the system.

In the present work, as particular examples and to simplify the resulting analysis, we use the infinite anchoring conditions discussed above. In particular, we choose two general anchoring conditions that are relevant to a variety of situations.
As explained below, we firstly consider patterned infinite anchoring, which we then specialise to the cases of unidirectional rubbed infinite anchoring with a constant pretilt and axisymmetric patterned infinite anchoring with a constant pretilt.
Secondly, we consider conical infinite anchoring, which we then specialise to the cases of homeotropic infinite anchoring and planar degenerate infinite anchoring.

\subsection{Patterned infinite anchoring}

For patterned infinite anchoring, the director has a fixed orientation at each location on each plate, where these directions could, in general, be different on the lower and upper plates. Specifically, patterned infinite anchoring on the lower and upper plates is given by
\begin{equation}
    \theta = \Theta_0(x,y) \quad \text{and} \quad \phi = \Phi_0(x,y) \quad \text{on} \quad z=0
\end{equation}
and
\begin{equation}
    \theta = \Theta_d(x,y) \quad \text{and} \quad \phi = \Phi_d(x,y) \quad \text{on} \quad z=d.
\end{equation}
In principle, this scenario allows for any patterned design to be considered on each plate; for example, an axisymmetric or periodic pattern, but here we restrict our attention to scenarios where the patterning is the same on both plates, so that
\begin{equation}
    \theta = \Theta(x,y) \quad \text{and} \quad \phi = \Phi(x,y) \quad \text{on} \quad z=0,\,d, \label{InfBC1}
\end{equation}
and consider the following two examples.

\subsubsection{Unidirectional rubbed infinite anchoring with a constant pretilt}

Unidirectional rubbed infinite anchoring may be achieved by coating the surfaces of the plates with a polymeric material and then mechanically rubbing the coating in a particular direction, called the rubbing direction, so that a single preferred director orientation is created \cite{SONINBOOK}. The directional rubbing process creates a preferred twist director angle $\Phir$, called the constant rubbing angle. This mechanical rubbing also often creates a preferred tilt director angle, called a pretilt angle, so that the director prefers to align at a fixed angle $\Thetat$ ($0 \le \Thetat \le \pi/2$) to the normal of the plates.
Here we assume that the preferred director orientation is the same on both plates, and so the boundary conditions for the director for unidirectional rubbed infinite anchoring with a constant pretilt on both the lower and upper plates are given by
\begin{equation}
    \theta = \Thetat \quad \text{and} \quad \phi=\Phir \quad \text{on} \quad z=0,\, d. \label{InfBC:rub}
\end{equation}

\subsubsection{Axisymmetric patterned infinite anchoring with a constant pretilt}

We also consider a case of non-uniform patterning, namely radially-independent, axisymmetric patterned infinite anchoring with a constant pretilt. Such non-uniform patterned anchoring may be achieved by optical or mechanical methods, for instance with patterned photo-alignment or ion-beam etching \cite{Chigrinov2021,NATUREpolyimide,HANDBOOK_ALIGNMENT,Hanaoka2004,Wu2007}, and here we consider the same patterned infinite anchoring with a constant pretilt on both the lower and upper plates, given by
\begin{equation}
    \theta = \Thetat \quad \text{and} \quad \phi=\Phic+\tan^{-1}\bigg(\dfrac{y}{x}\bigg) \quad \text{on} \quad z=0,\, d. \label{InfBC:axi}
\end{equation}
For the pattern given by \cref{InfBC:axi}, $\Phic$ is the twist angle between the director and the radial vector, and so when $\Phic = 0$ the projection of the director field onto the plates is a radial pattern, when $\Phic = \pi/2$ the projection of the director field onto the plates is an azimuthal (\ie a circular) pattern, and when $0<\Phic < \pi/2$ the projection of the director field onto the plates is a spiral pattern.

\subsection{Conical infinite anchoring}

For conical infinite anchoring, the director has a fixed pretilt angle to the plane of the plates, $\Thetat$, but is free to rotate about the normal of the plates and therefore lies on a cone with constant opening angle $\pi - 2\Thetat$. In this situation, the boundary condition on the twist director angle is derived from the fact that the torque on the director about the plate normal is zero \cite{ISBOOK}. For the Oseen–Frank bulk elastic energy \cref{WF0} this torque condition is $\phi_z=0$.
Conical infinite anchoring on the lower and upper plates is therefore given by
\begin{equation}
    \theta = \Thetat \quad \text{and} \quad \phi_z = 0 \quad \text{on} \quad z=0,\,d, \label{InfDBC}
\end{equation}
and we consider two extreme examples, namely homeotropic infinite anchoring and planar degenerate infinite anchoring.

\subsubsection{Homeotropic infinite anchoring}

The extreme case of conical infinite anchoring when $\Thetat=\pi/2$, so that the cone opening angle is zero, is termed homeotropic infinite anchoring, and occurs when the director has a preferred orientation perpendicular to the plates. Such anchoring is usually achieved through chemically treating the plates, for instance through coating with a surfactant such as lecithin \cite{SONINBOOK}. When the director is in this preferred orientation we see from \cref{director} that $\n =\unitz$ and the twist director angle $\phi$ is not defined at the plates. Homeotropic infinite anchoring on the lower and upper plates is then given by
\begin{equation}
    \theta = \dfrac{\pi}{2}\quad \text{and} \quad \phi_z = 0 \quad \text{on} \quad z=0,\, d. \label{InfBC:H}
\end{equation}

\subsubsection{Planar degenerate infinite anchoring}

The opposite extreme to homeotropic anchoring occurs when the cone opening angle is $\pi/2$ occurs when $\Theta_c=0$ and is known as planar degenerate infinite anchoring. In this case the director at each plate is parallel to the plane of the plate but is free to rotate around the normal to the plate.
Planar degenerate infinite anchoring on the lower and upper plates is then given by
\begin{equation}
    \theta = 0\quad \text{and} \quad \phi_z = 0 \quad \text{on} \quad z=0,\, d. \label{InfBC:PG}
\end{equation}

As mentioned above, many other types of anchoring are possible, several of which would lead to interesting situations, for example, weak anchoring conditions \cite{SONINBOOK}. A similar analysis to that described below may be possible, but to keep the resulting analysis analytically tractable, we will not pursue these other forms of anchoring in this work.

\section{Boundary Conditions: free surface}
\label{sec:BC:fs}

As we will see in \cref{sec:ER0,sec:ERinf}, the depth-averaged governing equations for nematic Hele-Shaw flow are formulated using only the boundary conditions for $\vel$ and $\n$ on the lower and upper plates. Boundary conditions for $\vel$, $\n$ and $p$ on the free surface $\partial \Omega$ may be subsequently needed in order to tackle specific situations in which, for example, the nematic is surrounded by an ambient gas, an isotropic fluid, a different nematic material, or a solid boundary. Such boundary conditions on the free surface can be derived through the usual approach of considering balances of mass, stress and torque \cite{Cousins2022,ISBOOK,SONINBOOK}. In the present work, we will not restrict ourselves to specific forms of the free surface boundary conditions in order to keep the approach as general as possible, until \cref{sec:app}, in which we consider the flow of nematic according to a simple model for the squeezing stage of the ODF method by prescribing the behaviour of the free surface.

We now introduce an appropriate nondimensionalisation before deriving the thin-film Ericksen--Leslie equations that govern the flow and director within a nematic Hele-Shaw cell.

\section{Nondimensionalisation}
\label{sec:ND}

We proceed by nondimensionalising all independent and dependent variables with appropriate scales. We assume that, because of whatever specific situation we are considering, we may define a characteristic lengthscale of variations in the $xy$-plane, which we denote by $L$, and a characteristic lengthscale of variations in the $z$-direction, which we denote by $\Hscale$. Lengths in the $xy$-plane are therefore nondimensionalised with $L$, while lengths in the $z$-direction are nondimensionalised with $D$, and we may define the nondimensional aspect ratio of the Hele-Shaw cell, denoted $\delta$, by
\begin{align}\label{delta}
    \delta = \dfrac{\Hscale}{L}.
\end{align}
The characteristic velocity scale in the $xy$-plane is denoted by $U$, for which there are several equally sensible choices, including:
$U = G\Hscale^2/\mu$ for a flow driven by a constant pressure gradient $G$;
$U = Q/(L\Hscale)$ for a flow driven by a prescribed flux $Q$; and
$U = \Hvscale\sqrt{V/(4\pi\Hscale^3)}$ for a flow driven by squeezing a circular cylindrical volume $V$ of nematic between parallel plates with a characteristic plate speed $\Hvscale$.
In what follows, we leave $U$ unspecified to keep the approach as general as possible for now, but in \cref{sec:app} we will use the velocity scale $U = \Hvscale\sqrt{V/(4\pi\Hscale^3)}$.
The characteristic timescale is denoted by $\tau$ and will also remain unspecified until later in this section, when the possible choices for $\tau$ will be described.
The conservation of mass equation \cref{CM} implies that the velocity scale in the $z$-direction is $\delta U$. The pressure is nondimensionalised so that it appears in the leading-order problem. Finally, all viscosities are nondimensionalised using the classical Newtonian viscosity $\mu = \eta_3 = \alpha_4/2$.
In summary, the Ericksen--Leslie equations \cref{LM1,LM2,LM3,CM,AM1,AM2} are nondimensionalised according to
\begin{align}\label{Scale}
\begin{gathered}
x = L \, x^*, \qquad y = L \, y^*, \qquad z = \Hscale z^* =\delta L \, z^*, \qquad t = \tau \, t^*, \qquad d = \Hscale d^* = \delta L d^*, \\
u = U \, u^*, \qquad v = U \, v^*, \qquad w = \delta U \, w^*, \qquad \tp = \dfrac{\mu U}{\delta^2 L} {\tp}^*, \\
\alpha_1 = \mu \, {\alpha_1}^*, \quad \alpha_2 = \mu \, {\alpha_2}^*, \quad \alpha_3 = \mu \, {\alpha_3}^*, \quad \alpha_4 = \mu \, {\alpha_4}^*, \quad \alpha_5 = \mu \, {\alpha_5}^*, \quad \alpha_6 = \mu \, {\alpha_6}^*, \\
\gamma_1 = \mu {\gamma_1}^*, \quad \gamma_2 = \mu {\gamma_2}^*, \quad \eta_1 = \mu \, {\eta_1}^*, \quad \eta_2 = \mu \, {\eta_2}^*, \quad \eta_3 = \mu \, {\eta_3}^*, \quad \eta_{12} = \mu \, {\eta_{12}}^*,
\end{gathered}
\end{align}
where the stars denote nondimensional variables. Henceforth, the stars are dropped, and all variables are nondimensional unless stated otherwise, except for the characteristic velocity scale $U$, and the characteristic timescale $\tau$.

\subsection{Thin-film flow}

We now proceed in a similar way to the standard thin-film approach used for isotropic Hele-Shaw flow, based on the assumption that the characteristic lengthscale of variations in the $xy$-plane is much larger than the characteristic lengthscale of variations in the $z$-direction, so that the aspect ratio $\delta=D/L$ is small, specifically $\delta\ll 1$.

After applying the nondimensionalisation \cref{Scale}, the conservation of mass equation \cref{CM} remains the same, and the linear momentum equations \cref{LM1,LM2,LM3} become
\begin{align}
    \dfrac{\tau_2}{\tau} \pdv{u}{t} + \delta \RE\bigg(u\pdv{u}{x}+v\pdv{u}{y}+w\pdv{u}{z}\bigg) =& -\pdv{\tp}{x} +\pdv{z} \Bigg[\gx(\theta,\phi)\pdv{u}{z} +\gxy(\theta,\phi)\pdv{v}{z} \Bigg] \nonumber \\
    & + \dfrac{\delta\tau_1}{\tau}\pdv{z} \Bigg[ m(\theta)\cos\phi\pdv{\theta}{t}+q(\theta)\sin\phi\pdv{\phi}{t} \Bigg]
    + \Oh\bigg(\delta,\delta^2\dfrac{\tau_1}{\tau}\bigg), \label{LM1eq} \\
    \dfrac{\tau_2}{\tau}\pdv{v}{t} + \delta \RE\bigg(u\pdv{v}{x}+v\pdv{v}{y}+w\pdv{v}{z}\bigg)=&-\pdv{\tp}{y}+\pdv{z} \Bigg[\gxy(\theta,\phi)\pdv{u}{z} +\gy(\theta,\phi)\pdv{v}{z} \Bigg] \nonumber \\
    & + \dfrac{\delta\tau_1}{\tau} \pdv{z} \Bigg[ m(\theta)\sin\phi\pdv{\theta}{t}-q(\theta)\cos\phi\pdv{\phi}{t} \Bigg]
    + \Oh\bigg(\delta,\delta^2\dfrac{\tau_1}{\tau}\bigg), \label{LM2eq} \\
    \dfrac{\delta^2\tau_2}{\tau} \pdv{w}{t} + \delta^3 \RE\bigg(u\pdv{w}{x}+v\pdv{w}{y}+w\pdv{w}{z}\bigg) =& - \pdv{\tp}{z} +\delta^2\dfrac{\tau_1}{\tau} \Bigg[\gamma_2 \pdv{z}\bigg(\sin\theta\cos\theta\pdv{\theta}{t}\bigg) - \gamma_1\bigg( \pdv{\theta}{z}\pdv{\theta}{t} +\cos^2\theta\pdv{\phi}{t}\pdv{\phi}{z}\bigg) \Bigg] \nonumber \\
    & + \Oh\bigg(\delta,\delta^3\dfrac{\tau_1}{\tau}\bigg), \label{LM3eq}
\end{align}
where $\gx(\theta,\phi)$, $\gy(\theta,\phi)$ and $\gxy(\theta,\phi)$ are effective viscosity functions defined by
\begin{align}
     \gx(\theta,\phi) &= \eta_1 \cos^2\theta\cos^2\phi+\eta_2\sin^2\theta+\cos^2\theta\sin^2\phi+\eta_{12}\sin^2\theta\cos^2\theta\cos^2\phi, \label{gx} \\
     \gy(\theta,\phi) &= \eta_1 \cos^2\theta\sin^2\phi+\eta_2\sin^2\theta+\cos^2\theta\cos^2\phi+\eta_{12}\sin^2\theta\cos^2\theta\sin^2\phi, \label{gy} \\
    \gxy(\theta,\phi) &= \eta_1 \cos^2\theta\sin\phi\cos\phi-\cos^2\theta\sin\phi\cos\phi+\eta_{12}\sin^2\theta \cos^2\theta\sin\phi\cos\phi. \label{gxy}
\end{align}
Note that the effective viscosity functions $\gx(\theta,\phi)$ and $\gy(\theta,\phi)$ are related through a $\pi/2$ shift in the twist angle $\phi$, \ie $\gx(\theta,\phi) = \gy(\theta,\pi/2-\phi)$, and that the effective viscosity functions can be related to the standard effective viscosity functions $g(\theta)$ and $h(\theta)$ \cite{ISBOOK}, which are defined by
\begin{align}
    g(\theta) &= \eta_1 \cos^2 \theta + \eta_2 \sin^2 \theta +\eta_{12}\sin^2\theta\cos^2\theta, \label{gfunc} \\
    h(\theta) &= \eta_2 \sin^2 \theta+\cos^2\theta, \label{hfunc}
\end{align}
according to
\begin{align}
    h(\theta) &= \gx(\theta,\pi/2) = \gy(\theta,0), \\
    g(\theta) &= \gx(\theta,0) =\gy(\theta,\pi/2) =\gx(\theta,\phi) +\gxy(\theta,\phi) \, \tan\phi=\gy(\theta,\phi) + \gxy(\theta,\phi) \, \cot\phi, \\
    g(\theta)\,h(\theta) &=\gx(\theta,\phi)\,\gy(\theta,\phi)-\gxy(\theta,\phi)^2.
\end{align}
Also appearing in \cref{LM1eq,LM2eq,LM3eq} are two timescales, namely the timescale on which fluid travels the length of the cell, $\tau_1$, which is defined as
\begin{align}
    \tau_1 &= \dfrac{L}{U}, \label{tau2}
\end{align}
and the fluid inertia timescale $\tau_2$, which is defined in terms of the reduced Reynolds number $\delta \RE$ and $\tau_1$ as
\begin{align}
    \tau_2 &= \delta \, \RE \, \tau_1 = \dfrac{\rho \Hscale^2}{\mu}, \label{tau3}
\end{align}
where the usual Reynolds number $\RE$, which measures the ratio of inertial effects to viscous effects within the system, is defined by
\begin{align}
    \RE &= \dfrac{\rho U \Hscale}{\mu}. \label{Reynolds}
\end{align}
As usual, a large Reynolds number therefore corresponds to the situation in which inertial effects are much stronger than viscous effects, while a small Reynolds number corresponds to the opposite situation in which viscous effects are much stronger than inertial effects.

After nondimensionalisation, the no-slip and no-penetration conditions \cref{noslipH,noslip0} are given by
\begin{equation}
     u = v = w = 0 \quad \text{on} \quad z = 0 \label{noslip0nd}
\end{equation}
and
\begin{equation}
    u = v = 0 \quad \text{and} \quad w = \dfrac{\tau_1}{\tau} d' \quad \text{on} \quad z = d. \label{noslipHnd}
\end{equation}

Similarly, applying the nondimensionalisation \cref{Scale} to the angular momentum equations \cref{AM1,AM2}, and collecting terms in orders of $\delta$ yields
\begin{align}\label{AM1eq}
\begin{gathered}
     \gamma_1 \, \dfrac{\tau_3}{\tau} \pdv{\theta}{t} = \pdv[2]{\theta}{z}+\sin\theta\cos\theta \Big(\pdv{\phi}{z}\Big)^2 +\delta^2 \Bigg[ \pdv[2]{\theta}{x} +\pdv[2]{\theta}{y} + \sin\theta\cos\theta \bigg( \Big(\pdv{\phi}{x}\Big)^2+\Big(\pdv{\phi}{y}\Big)^2 \bigg) \Bigg] \\
     \mbox{} - \ER \, m(\theta) \bigg[\cos\phi \, \pdv{u}{z} + \sin\phi \, \pdv{v}{z} \bigg] +\dfrac{\delta^2 L^2}{K}\pdv{\hPsi}{\theta}+\Oh(\delta\ER),
\end{gathered}
\end{align}
and
\begin{align}\label{AM2eq}
\begin{gathered}
     \gamma_1 \, \dfrac{\tau_3}{\tau} \cos^2\theta \pdv{\phi}{t} = \pdv{z} \Bigg( \cos^2 \theta \pdv{\phi}{z} \Bigg) + \delta^2 \Bigg[ \pdv{x}\bigg(\cos^2 \theta \pdv{\phi}{x}\bigg) +\pdv{y}\bigg(\cos^2 \theta \pdv{\phi}{y}\bigg) \Bigg] \\
     \mbox{} - \ER \, q(\theta) \bigg(\sin \phi \,\pdv{u}{z} -\cos \phi \,\pdv{v}{z} \bigg) +\dfrac{\delta^2 L^2}{K}\pdv{\hPsi}{\phi}+\Oh(\delta\ER),
\end{gathered}
\end{align}
where $\ER$ is the Ericksen number, defined by
\begin{align}
    \ER = \dfrac{\mu \Hscale U}{K}, \label{Ericksen}
\end{align}
is a measure of the ratio of viscous effects to elasticity effects within the system.
A large Ericksen number therefore corresponds to the situation in which viscous effects are much stronger than elasticity effects, while a small Ericksen number corresponds to the opposite situation in which elasticity effects are much stronger than viscous effects.
Also appearing in \cref{AM1eq,AM2eq} is a third timescale, the director rotation timescale $\tau_3$, which is defined as
\begin{align}
    \tau_3 &= \dfrac{\mu \Hscale^2}{K}. \label{tau1}
\end{align}
Additionally, the functions $m(\theta)$ and $q(\theta)$ appearing in \cref{AM1eq,AM2eq} are effective viscosity functions, which are defined by
\begin{align}
    m(\theta) = \dfrac{1}{2} \, (\gamma_1 + \gamma_2 \cos2\theta)\quad \text{and}\quad
    q(\theta) = \dfrac{1}{2} \, (\gamma_1-\gamma_2)\sin\theta\cos\theta. \label{mqfunc}
\end{align}
The anchoring conditions on the plates discussed in \cref{sec:BC}, are unchanged after applying the nondimensionalisation \cref{Scale}.

Note that in \cref{AM1eq,AM2eq,LM1eq,LM2eq,LM3eq} we have retained the terms with coefficients of $\tau_1/\tau$, $\tau_2/\tau$ and $\tau_3/\tau$ that are the lowest order in $\delta$ as the choice of the timescale $\tau$ discussed subsequently may change the order in $\delta$ at which these terms appear.

\subsection{Characteristic timescales}
\label{sec:thinfilm:tscales}

\begin{table}[tp]
\begin{center}
{\renewcommand{\arraystretch}{1}
\begin{tabular}{|cc|c|c|c|c|c|c|}
    \hline
    && $\quad\:\tau_1\,[{\rm s}]\quad\:$ & $\:\quad\tau_2\,[{\rm s}]\quad\:$ & $\quad\:\tau_3\,[{\rm s}]\quad\:$ & $\:\qquad\delta\qquad\:$ & $\qquad\RE\qquad$ & $\qquad\ER\qquad$ \\
    \hline
    \quad Analysis of the ODF method \quad & min. & $5.0\times10^{-2}$ & $2.5\times10^{-8}$ & $2.5\times10^{-4}$ & \multirow{2}{*}{$1.0\times10^{-2}$} & $1.0\times10^{-5}$ & $1.0\times10^{-1}$ \\
    by Cousins \etal \cite{Cousins2019} & max. & $2.5\times10^{-1}$ & $2.5\times10^{-4}$ & $2.5$ & & $5.0\times10^{-1}$ & $5.0\times10^{3}$ \\
    \hline
    \multicolumn{2}{|c|}{Capillary-filling experiments} & \multirow{2}{*}{$1.0\times 10^{3}$} & \multirow{2}{*}{$3.1\times 10^{-6}$} & \multirow{2}{*}{$4.5\times 10^{-1}$} & \multirow{2}{*}{$1.0\times10^{-4}$} & \multirow{2}{*}{$3.0\times10^{-5}$} & \multirow{2}{*}{$4.5$} \\
    \multicolumn{2}{|c|}{by Mi and Yang \cite{CapillaryMiYang}} &&&&&& \\
    \hline
    \multicolumn{2}{|c|}{Viscous fingering experiments} & \multirow{2}{*}{$1.2\times 10^{-2}$} & \multirow{2}{*}{$3.8\times 10^{-7}$} & \multirow{2}{*}{$5.5\times 10^{-2}$} & \multirow{2}{*}{$3.5\times10^{-2}$} & \multirow{2}{*}{$8.8\times10^{-4}$} & \multirow{2}{*}{$1.3\times10^{2}$} \\
    \multicolumn{2}{|c|}{by Sonin and Bartolino \cite{Sonin1993}} &&&&&& \\
    \hline
    \quad Nematic microfluidic experiments \quad & min. & $3.0\times10^{1}$ & $7.8\times10^{-5}$ & $2.9\times10^{-1}$ & $4.0\times10^{-4}$ & $5.0\times10^{-5}$ & $7.2$ \\
    by Sengupta \etal \cite{Sengupta2013d} & max. & $1.0\times10^{2}$ & $2.0\times10^{-6}$ & $1.1\times10^{1}$ & $2.5\times10^{-3}$ & $1.0\times10^{-3}$& $1.5\times10^{2}$ \\
    \hline
\end{tabular}}
\caption{
Typical values for the timescales $\tau_1$, $\tau_2$ and $\tau_3$ and the nondimensional numbers $\delta$, $\RE$ and $\ER$ for four different situations, namely analysis of the ODF method by Cousins \etal \cite{Cousins2018,Cousins2019}, capillary-filling experiments by Mi and Yang \cite{CapillaryMiYang}, viscous fingering experiments by Sonin and Bartolino \cite{Sonin1993}, and nematic microfluidic experiments by Sengupta \etal \cite{Sengupta2013d}.
A full statement of all of the parameter values used to generate these values is given in \cref{sec:appendixb}.
}
\label{table:examples}
\end{center}
\end{table}

An inspection of the nondimensional Ericksen--Leslie equations \cref{AM1eq,AM2eq,LM1eq,LM2eq,LM3eq,CM} shows that there are three natural choices for the timescale $\tau$. In situations in which time-dependent changes in the director angles are important, it would be appropriate to nondimensionalise time with the director rotation timescale $\tau_3$, so that $\tau = \tau_3$.
This choice of timescale may be appropriate, for instance, when modelling the director rotation due to flow within a channel for which the plates exhibit homeotropic anchoring \cite{Sengupta2013} (for one-dimensional models of these transitions, see Anderson \etal \cite{Anderson2015} and Crespo \etal \cite{Crespo2017}).
In situations in which the dynamics of the flow across the lengthscale $L$ are of interest, it would be natural to use the timescale $\tau = \tau_1$.
On the other hand, when dynamics induced by inertial effects are of particular interest, the choice of timescale $\tau = \tau_2$ is appropriate.
\cref{table:examples} shows typical values for the timescales $\tau_1$, $\tau_2$ and $\tau_3$ and the nondimensional numbers $\delta$, $\RE$ and $\ER$ for four different situations, namely analysis of the ODF method by Cousins \etal \cite{Cousins2019}, capillary-filling experiments by Mi and Yang \cite{CapillaryMiYang}, viscous fingering experiments by Sonin and Bartolino \cite{Sonin1993}, and nematic microfluidic experiments by Sengupta \etal \cite{Sengupta2013d}. From \cref{table:examples} we see that, in all but one extreme case, we have $\tau_1 \gg \tau_2,\tau_3$. There is a similar situation for many situations of relevance to LCD manufacturing \cite{Cousins2019,Cousins2020}, viscous fingering experiments \cite{Sonin1993,Lam1989,TothKatona2003,Folch2000}, and some experiments with nematic microfluidic channels \cite{Sengupta2011,Sengupta2012,Sengupta2013}.

Given the many applications for which flow over the lengthscale in the $xy$-plane is relevant, \eg the cases mentioned above for LCD manufacturing, viscous fingering experiments, and nematic microfluidic experiments, here we choose the timescale to be $\tau_1$.
Specifically, we set $\tau = \tau_1$ and therefore
\begin{align}
    \dfrac{\tau_1}{\tau} = 1, \quad \dfrac{\tau_2}{\tau} = \delta \, \RE \quad \text{and} \quad \dfrac{\tau_3}{\tau} = \delta \, \ER.
\end{align}
Additionally, we assume that viscous effects are much stronger than inertial effects, and hence we assume that the reduced Reynolds number $\delta \,\RE$ is small, such that $\delta\Re \ll 1$.
This is certainly the case for examples given in \cref{table:examples}, where $\delta \, \RE \approx 10^{-7}$--$10^{-9}$.
Also, as mentioned previously, we neglect any conservative body forces and hence set $\hPsi \equiv 0$.

\section{Thin-film Ericksen--Leslie equations}
\label{sec:thinfilm}

We now proceed by employing the standard thin-film approach used for isotropic Hele-Shaw flow and consider only the leading-order problem in the limit $\delta \to 0$.
In this limit, and setting $\tau = \tau_1$, $\delta\Re \ll 1$ and $\hPsi \equiv 0$ in \cref{LM1eq,LM2eq,LM3eq,CM,AM1eq,AM2eq}, as discussed above, the leading-order Ericksen--Leslie equations, hereafter referred to as the thin-film Ericksen--Leslie equations, are given by
\begin{align}
    0 &= \pdv{u}{x} +\pdv{v}{y} +\pdv{w}{z}, \label{CMeq0} \\
    \pdv{\tp}{x}&= \pdv{z} \Bigg[\gx(\theta,\phi)\pdv{u}{z} +\gxy(\theta,\phi)\pdv{v}{z} \Bigg], \label{LM1eq0} \\
    \pdv{\tp}{y}&=\pdv{z} \Bigg[\gxy(\theta,\phi)\pdv{u}{z} +\gy(\theta,\phi)\pdv{v}{z}\Bigg] , \label{LM2eq0} \\
    \pdv{\tp}{z}&=0, \label{LM3eq0} \\
    0&= \pdv[2]{\theta}{z}+\sin\theta\cos\theta \Big(\pdv{\phi}{z}\Big)^2 - {\ER} \, m(\theta) \bigg[\cos\phi \, \pdv{u}{z} + \sin\phi \, \pdv{v}{z} \bigg], \label{AM1eq0} \\
    0&= \pdv{z} \Bigg[ \cos^2 \theta \pdv{\phi}{z} \Bigg]- \ER \, q(\theta) \bigg(\sin \phi \,\pdv{u}{z} -\cos \phi \,\pdv{v}{z} \bigg). \label{AM2eq0}
\end{align}
Although \cref{LM1eq0,LM2eq0,LM3eq0,CMeq0,AM1eq0,AM2eq0} do not include any time derivatives, we note that their solutions can still depend on time $t$ in situations in which the boundary conditions on the plates and/or the free surface are time dependent.

Equation \cref{LM3eq0} shows that the pressure is independent of $z$, and hence $\tp=\tp(x,y,t)$, but solving the remaining thin-film Ericksen--Leslie equations \cref{LM1eq0,LM2eq0,CMeq0,AM1eq0,AM2eq0} is, in general, difficult and may require a numerical approach.
In the present work, we take an alternative approach and analyse the thin-film Ericksen--Leslie equations in a number of limiting cases in which we can make significant analytical progress.
Firstly, in \cref{sec:ER0}, we consider the leading-order problem in the limiting case in which elasticity effects are much stronger than viscous effects, and hence the Ericksen number is small ($\ER \ll 1$).
Examples of such situations include the capillary-filling method where flow is driven by capillary action \cite{CapillaryMiYang} and flows driven by gravity \cite{LamCummings2015}.
Secondly, in \cref{sec:ERinf}, we consider the leading-order problem in the limiting case in which viscous effects are much stronger than viscous effects, and hence the Ericksen number is large ($\ER \gg 1$).
Examples of such situations include the ODF method where flow is driven by squeezing \cite{Cousins2019,Cousins2020} and in recent experiments using nematic microfluidic devices \cite{Sengupta2013,Sengupta2013b,Zhao2022,Stieger2017} in which the flow is driven by a large pressure gradient.

\section{The limit of small Ericksen number ($\ER \ll 1$)}
\label{sec:ER0}

In this section, we consider the leading-order problem in the limit of small Ericksen number ($\ER\ll1$), with all the scenarios of anchoring mentioned in \cref{sec:BC}. We consider the general case of patterned infinite anchoring in \cref{sec:ER0:pat}, with particular cases of unidirectional rubbed infinite anchoring with a constant pretilt and axisymmetric patterned infinite anchoring with a constant pretilt, and then consider the general case of conical infinite anchoring in \cref{sec:ER0:con}, with particular cases of homeotropic infinite anchoring and planar degenerate infinite anchoring.

\subsection{Patterned infinite anchoring}
\label{sec:ER0:pat}

We begin by considering the scenario of patterned infinite anchoring, which corresponds to the anchoring conditions \cref{InfBC1}.
At leading order in $\ER\ll 1$, the thin-film conservation of angular momentum equations \cref{AM1eq0,AM2eq0} subject to \cref{InfBC1} are satisfied by the director angle solutions
\begin{align}
    \phi = \Phi(x,y) \quad \text{and} \quad \theta = \Theta(x,y).
\end{align}
Therefore, in this limit, the director field throughout the cell is identical to the director field patterned on the plates.
We note that in the scenario where the patterned infinite anchoring on the two plates is different, a numerical approach is, in general, required to solve for the leading-order director angles $\theta$ and $\phi$.

At leading order in $\ER \ll 1$, the thin-film conservation of linear momentum equations \cref{LM1eq0,LM2eq0} are given by
\begin{align}
    \pdv{\tp}{x} = \mgx \, \pdv[2]{u}{z} +\mgxy \, \pdv[2]{v}{z} \quad \text{and} \quad
    \pdv{\tp}{y} = \mgxy \, \pdv[2]{u}{z} +\mgy \, \pdv[2]{v}{z}, \label{LM12eq1}
\end{align}
where $\mgx = \gx(\Phi(x,y),\Theta(x,y))$, $\mgy = \gy(\Phi(x,y),\Theta(x,y))$ and $\mgxy = \gxy(\Phi(x,y),\Theta(x,y))$.
Integrating \cref{LM12eq1} with respect to $z$ twice, applying the no-slip conditions \cref{noslip0nd,noslipHnd}, and rearranging yields solutions for $u$ and $v$
\begin{align}
    u = \dfrac{1}{2\mg\mh} \left(\mgy\pdv{\tp}{x} - \mgxy \pdv{\tp}{y}\right)z (z-d) \quad \text{and} \quad
    v = \dfrac{1}{2\mg\mh} \left(\mgx\pdv{\tp}{y} - \mgxy \pdv{\tp}{x}\right)z (z-d), \label{uvLowERinf}
\end{align}
where $\mg=g(\Theta(x,y))$ and $\mh=h(\Theta(x,y))$.
The patterned anchoring, therefore, creates a fixed director field which in turn produces an anisotropic patterned viscosity via the effective viscosity functions $\mg$, $\mh$, $\mgx$, $\mgy$ and $\mgxy$.
The flow is then driven by the pressure gradients $\partial \tp / \partial x$ and $\partial \tp / \partial y$ and differs from the simple isotropic situation due to the patterned viscosity. The streamlines of the flow may therefore be tailored by using plates on which patterned anchoring has been created. This tailoring of the streamlines is an example of the flow being guided by the director field, a situation that has previously been investigated theoretically by Leslie \cite{Leslie1979}, albeit only for a unidirectional director field. However, in this analysis, the fixed director was induced by a strong magnetic field and not by anchoring.

Following the standard approach used in the analysis of Hele-Shaw flow, we now substitute the solutions for the velocity \cref{uvLowERinf} into the conservation of mass equation \cref{CMeq0}, integrate with respect to $z$ between $z=0$ and $z=d$, and apply the no-slip and no-penetration conditions \cref{noslip0nd,noslipHnd}, to give the governing equation for the pressure $\tp$, namely
\begin{align}
    \pdv{x}\bigg( \dfrac{1}{\mg \mh} \left(\mgy \pdv{\tp}{x} - \mgxy\pdv{\tp}{y}\right) \bigg) + \pdv{y}\bigg( \dfrac{1}{\mg \mh} \left(\mgx \pdv{\tp}{y} - \mgxy\pdv{\tp}{x}\right) \bigg) = \dfrac{12d'}{d^3}. \label{pLowERinf}
\end{align}
Finally, we repeat this process by substituting the solutions for the velocity \cref{uvLowERinf} into the conservation of mass equation \cref{CMeq0}, but now integrating with respect to $z$ between $z=0$ and $z$, applying the no-slip condition \cref{noslip0nd}, and simplifying the expression by substituting \cref{pLowERinf} to give the vertical velocity,
\begin{align}
    w = \dfrac{d'}{d^3} \, (3d-2z)\, z^2, \label{wLowERinf}
\end{align}
which is independent of the director angles and identically zero when the upper plate is stationary.
After $\tp$ has been obtained from \cref{pLowERinf}, the velocity components $u$ and $v$ can be calculated from \cref{uvLowERinf}.
In general, for a non-homogeneous anchoring pattern, the solution for $\tp$ from \cref{pLowERinf} must be obtained numerically; however, as we shall see shortly, there are cases in which the symmetry of the anchoring pattern allows for further analytical progress.

In summary, in the scenario of patterned infinite anchoring, the thin-film Ericksen--Leslie equations \cref{LM1eq0,LM2eq0,LM3eq0,CMeq0,AM1eq0,AM2eq0} can be written in terms of the unknown pressure $\tp$ as
\begin{align}\label{ER0:inf}
\begin{gathered}
    u  = \dfrac{1}{2\mg \mh} \left(\mgy \pdv{\tp}{x} -\mgxy \pdv{\tp}{y}\right)z (z-d), \quad
    v = \dfrac{1}{2\mg \mh} \left(\mgx\pdv{\tp}{y} -\mgxy \pdv{\tp}{x}\right)z (z-d), \quad
    w = \dfrac{d'}{d^3} \, (3d-2z)\, z^2, \\
    \pdv{x}\bigg( \dfrac{1}{\mg \mh} \left(\mgy \pdv{\tp}{x} -\mgxy\pdv{\tp}{y}\right) \bigg) + \pdv{y}\bigg( \dfrac{1}{\mh \mg} \left(\mgx \pdv{\tp}{y} -\mgxy\pdv{\tp}{x}\right) \bigg) = \dfrac{12d'}{d^3}, \quad
    \theta \equiv \Theta(x,y), \quad
    \phi = \Phi(x,y).
\end{gathered}
\end{align}

We now consider the particular cases of unidirectional rubbed infinite anchoring with a constant pretilt and axisymmetric patterned infinite anchoring with a constant pretilt.

\subsubsection{Unidirectional rubbed infinite anchoring with a constant pretilt}

For unidirectional rubbed infinite anchoring with a constant pretilt, namely \cref{InfBC1} with \cref{InfBC:rub}, the solution to \cref{ER0:inf} may be obtained using a rotation of the $xy$-coordinate system to a new $\hat{x}\hat{y}$-coordinate system in which the projection of the preferred director at the plates is a coordinate axis,
\begin{align} \label{trans2}
    \xr = \dfrac{1}{ \sqrt{\hc} } \big(  \cos\Phir \, x + \sin\Phir \, y \big), \qquad
    \yr = \dfrac{1}{ \sqrt{\gc} } \big( -\sin\Phir \, x + \cos\Phir \, y \big),
\end{align}
where $\hc = h(\Thetat)$ and $\gc = g(\Thetat)$.
At leading order in $\ER \ll 1$, the thin-film Ericksen--Leslie equations \cref{LM1eq0,LM2eq0,LM3eq0,CMeq0,AM1eq0,AM2eq0} can then be written in terms of the unknown pressure $\tp$ as
\begin{align}\label{ER0:inf:rub}
\begin{gathered}
    \hat{u} = \dfrac{1}{2 \gc} \pdv{\tp}{\xr} z (z-d), \quad
    \hat{v} = \dfrac{1}{2 \hc} \pdv{\tp}{\yr} z (z-d), \quad
    w = \dfrac{d'}{d^3} \, (3d-2z)\, z^2, \\
    \pdv[2]{\tp}{\xr} + \pdv[2]{\tp}{\yr} = \dfrac{12 \hc \gc d'}{d^3}, \quad
    \theta \equiv \Thetat, \quad \phi \equiv \Phir.
\end{gathered}
\end{align}
In \cref{ER0:inf:rub}, $\hat{u}$ and $\hat{v}$ are the velocity components parallel and perpendicular to the rubbing direction, respectively.
Note that $\hat{u}$ and $\hat{v}$ given by \cref{ER0:inf:rub} can be reformulated in terms of the gradient of the pressure in the original Cartesian coordinates as
\begin{align}
    \hat{u} = \sqrt{\dfrac{\hc}{4 \gc^2}} \bigg( \cos\Phir \pdv{\tp}{x} + \sin \Phir \pdv{\tp}{y} \bigg) z (z-d) \quad \text{and} \quad
    \hat{v} = \sqrt{\dfrac{\gc}{4 \hc^2}} \bigg( -\sin\Phir \pdv{\tp}{x} + \cos \Phir \pdv{\tp}{y} \bigg) z (z-d). \label{ER0:inf:rub:uvr}
\end{align}

For instance, a constant pressure gradient applied in the $x$-direction, \ie when $G = \partial \tp/\partial x$ and $\partial \tp/\partial y=0$, leads to a flow given by
\begin{align}\label{ER0:inf:rub:uvr2}
    \hat{u} =   \sqrt{\dfrac{\hc}{4 \gc^2}} \, \cos\Phir \, G z (z-d) \quad \text{and} \quad
    \hat{v} = - \sqrt{\dfrac{\gc}{4 \hc^2}} \, \sin\Phir \, G z (z-d).
\end{align}
The solutions in \cref{ER0:inf:rub:uvr2} show that the flow is driven by the pressure gradient in the $x$-direction but guided by the patterned viscosity that has been induced by the rubbed anchoring.
In particular, \cref{ER0:inf:rub:uvr2} shows that the magnitude of the velocity component parallel to the rubbing direction $|\hat{u}|$ is greater than the magnitude of the velocity component perpendicular to the rubbing direction $|\hat{v}|$ provided that $\tan \Phir<(\hc/\gc)^{3/2}$.
So, for example, for unidirectional rubbed anchoring with $\Phir = \pi/4$ and $\Thetat = 0$, and for the nematic 4'-pentyl-4-biphenylcarbonitrile (5CB) \cite{E7Mixture}, which has dimensional viscosity values $\eta_1=0.0204$\,Pa\,s and $\eta_3=0.0326$\,Pa\,s, and therefore nondimensional viscosity value ${\eta_1}^*=\eta_1/\eta_3=0.626$, we have $\tan\Phir = 1$ and $(\hc/\gc)^{3/2} = (1/{\eta_1}^*)^{3/2} = 2.02$, and hence $|\hat{u}|/|\hat{v}|=2.02$ and the flow is predominately in the rubbing direction.

\subsubsection{ Axisymmetric infinite anchoring with a constant pretilt}

For axisymmetric infinite anchoring with a constant pretilt, namely \cref{InfBC1} with \cref{InfBC:axi}, we use the polar coordinate transform,
\begin{align} \label{trans1}
    x = r \cos (\beta - \Phi_{\rm c}), \quad y = r \sin(\beta - \Phi_{\rm c})
\end{align}
in \cref{ER0:inf}, where $r$ and $\beta$ are the usual radial and azimuthal coordinates, respectively, and at leading order in $\ER \ll 1$ the thin-film Ericksen--Leslie equations \cref{LM1eq0,LM2eq0,LM3eq0,CMeq0,AM1eq0,AM2eq0} can be written in terms of the unknown $\tp$ as
\begin{align}\label{ER0:inf:axi}
\begin{gathered}
    u_r     = \dfrac{1}{2\cg \, \ch} \left(\cgy \pdv{\tp}{r} + 2\cgxy \dfrac{1}{r} \pdv{\tp}{\beta}\right)z (z-d), \quad
    u_\beta = \dfrac{1}{2\cg \, \ch} \left(\cgx \dfrac{1}{r} \pdv{\tp}{\beta} + 2\cgxy \pdv{\tp}{r}\right)z (z-d), \quad
    w = \dfrac{d'}{d^3} \, (3d-2z)\, z^2, \\
    \cgx \dfrac{1}{r^2} \pdv[2]{\tp}{\beta} + 2 \cgxy \dfrac{1}{r} \pdv{\tp}{\beta}{r} + \cgy \dfrac{1}{r} \pdv{r} \bigg (r \pdv{\tp}{r}\bigg) = \dfrac{12d' \, \cg \,\ch}{d^3}, \quad
    \theta \equiv \Thetat, \quad
    \phi = \Phic+\beta,
\end{gathered}
\end{align}
where $\cgx = \gx(\Thetat, \Phi_{\rm c})$, $\cgy = \gy(\Thetat, \Phi_{\rm c})$ and $\cgxy = \gxy(\Thetat, \Phi_{\rm c})$.

Inspection of \cref{ER0:inf:axi,gxy} shows that in situations in which the anchoring pattern is strictly radial or strictly azimuthal, \ie when $\Phic = 0$ or $\Phic = \pi/2$, then $\cgxy \equiv 0$ and there is a radial-flow solution that satisfies \cref{ER0:inf:axi} for which $\partial \tp/\partial \beta = 0$ and hence $u_\beta \equiv 0$.
In these situations, $\tp$ can be obtained by direct integration of the pressure equation in \cref{ER0:inf:axi} subject to appropriate boundary conditions on $\partial \Omega$.
Conversely, in situations in which the anchoring pattern is not strictly radial or strictly azimuthal, \ie when $\Phic \neq 0$ or $\Phic \neq \pi/2$, then $\cgxy \neq 0$ and no purely radial-flow solution satisfies \cref{ER0:inf:axi}, and the flow is a spiral, guided by the axisymmetric anchoring pattern.

\subsection{Conical infinite anchoring}
\label{sec:ER0:con}

For conical infinite anchoring, which corresponds to the anchoring conditions \cref{InfDBC}, at leading order in $\ER\ll1$,
the thin-film conservation of angular momentum equations \cref{AM1eq0,AM2eq0} are satisfied by the director angle solutions
\begin{align} \label{conic:sol}
    \theta \equiv \Thetat \quad \text{and} \quad \phi = \phi (x,y,t).
\end{align}
To determine $\phi$ we must consider higher-order thin-film conservation of angular momentum equations (specifically, \cref{AM1eq,AM2eq} at first-order in $\delta^2$).
Provided that $\ER \ll \delta^2 \ll 1$, we find that equations yield that the twist director angle $\phi$ is governed by Laplace's equation, namely
\begin{align}
   \pdv[2]{\phi}{x} + \pdv[2]{\phi}{y} = 0, \label{phiLaplace}
\end{align}
subject to appropriate boundary conditions on $\partial \Omega$.
Note that unlike in the scenario of patterned anchoring, previously discussed in \cref{sec:ER0:pat}, in which the anchoring on the plates fixes the director field throughout the cell, in the scenario of conical anchoring, the director field is determined by the anchoring on both the plates and $\partial\Omega$.

At leading order in $\ER\ll1$, the thin-film Ericksen--Leslie equations \cref{LM1eq0,LM2eq0,LM3eq0,CMeq0,AM1eq0,AM2eq0} can therefore be written in terms of the unknown pressure $\tp$ and twist angle $\phi$ as
\begin{align}\label{ER0:inf:conical}
\begin{gathered}
    u = \dfrac{1}{2\hc \gc} \left(\gy\pdv{\tp}{x} -\gxy \pdv{\tp}{y}\right)z (z-d), \quad
    v = \dfrac{1}{2\hc \gc} \left(\gx\pdv{\tp}{y} -\gxy \pdv{\tp}{x}\right)z (z-d), \quad
    w = \dfrac{d'}{d^3} \, (3d-2z)\, z^2 \\
    \pdv{x}\bigg( \gy \pdv{\tp}{x} -\gxy\pdv{\tp}{y} \bigg) + \pdv{y}\bigg(\gx \pdv{\tp}{y} -\gxy\pdv{\tp}{x} \bigg) = \dfrac{12\hc \gc d'}{d^3}, \quad
    \theta \equiv \Thetat, \quad
    \pdv[2]{\phi}{x} + \pdv[2]{\phi}{y} = 0.
\end{gathered}
\end{align}
We now consider the particular cases of homeotropic infinite anchoring and planar degenerate infinite anchoring.

\subsubsection{Homeotropic infinite anchoring}
\label{sec:ER0:conH}

For homeotropic infinite anchoring, which corresponds to the anchoring conditions \cref{InfBC1} with $\Theta = \pi/2$, at leading order in $\ER \ll 1$ the thin-film Ericksen--Leslie equations \cref{LM1eq0,LM2eq0,LM3eq0,CMeq0,AM1eq0,AM2eq0} can be written in terms of the unknown pressure $\tp$ as
\begin{align}\label{ER0:inf:H}
\begin{gathered}
    u = \dfrac{1}{2\eta_2} \pdv{\tp}{x} z (z-d), \quad
    v = \dfrac{1}{2\eta_2} \pdv{\tp}{y} z (z-d), \quad
    w = \dfrac{d'}{d^3} \, (3d-2z)\, z^2, \\
    \pdv[2]{\tp}{x} + \pdv[2]{\tp}{y} = \dfrac{12 \eta_2 d'}{d^3}, \quad
    \theta \equiv \dfrac{\pi}{2},
\end{gathered}
\end{align}
and we note that with $\theta \equiv \pi/2$ the twist director angle $\phi$ is not defined. In this situation, the director is therefore fixed perpendicular to the $xy$-plane, \ie $\n = \unitz$, throughout the cell (a situation sometimes called uniform homeotropic orientation).
The governing equation for the pressure, given in \cref{ER0:inf:H}, takes the same form as for isotropic Hele-Shaw flow, and therefore, the flow of a nematic with homeotropic infinite anchoring is identical to that of the flow of an isotropic fluid with viscosity $\eta_2$.

\subsubsection{Planar degenerate infinite anchoring}
\label{sec:ER0:conP}

\begin{table}[tp]
\centering
\setlength\tabcolsep{0.1cm}
\renewcommand{\arraystretch}{1.75}
\begin{tabular}{|ccccccc|}
    \hline
    & \multicolumn{6}{|c|}{Small Ericksen number $\ER \ll 1$} \\
    \hline
    & \multicolumn{3}{|c|}{Patterned infinite anchoring (\cref{sec:ER0:pat})} & \multicolumn{3}{c|}{Conical infinite anchoring (\cref{sec:ER0:con})} \\
    \hline
    & \multicolumn{1}{|c|}{General} & \multicolumn{1}{c|}{Rubbed} & \multicolumn{1}{c|}{Axisymmetric} & \multicolumn{1}{c|}{General} & \multicolumn{1}{c|}{Homeotropic} & \multicolumn{1}{c|}{Planar degenerate} \\
    \hline $\theta$ & \multicolumn{1}{|c|}{$\Theta(x,y)$} & \multicolumn{3}{c|}{$\Thetat$} & \multicolumn{1}{c|}{$\dfrac{\pi}{2}$} & $0$ \\
    \hline
    $\phi$ & \multicolumn{1}{|c|}{$\Phi(x,y)$} & \multicolumn{1}{c|}{$\Phic$} & \multicolumn{1}{c|}{$\Phic+\beta$} & \multicolumn{1}{c|}{$\pdv[2]{\phi}{x}+\pdv[2]{\phi}{y} =0$} & \multicolumn{1}{c|}{undefined} & \multicolumn{1}{c|}{$\pdv[2]{\phi}{x}+\pdv[2]{\phi}{y} =0$} \\
    \hline
    $\tp$ & \multicolumn{1}{|c|}{\cref{pLowERinf}} & \multicolumn{1}{c|}{$\pdv[2]{\tp}{\xr}+\pdv[2]{\tp}{\yr}= \dfrac{12 \hc \gc d'}{d^3}$} & \multicolumn{1}{c|}{\cref{ER0:inf:axi}} & \multicolumn{1}{c|}{\cref{ER0:inf:conical}} & \multicolumn{1}{c|}{$\pdv[2]{\tp}{x}+\pdv[2]{\tp}{y}= \dfrac{12 \eta_2 d'}{d^3}$} & \multicolumn{1}{c|}{\cref{ER0:inf:pland}} \\
    \hline
    $u$ & \multicolumn{1}{|c|}{\cref{uvLowERinf}} & \multicolumn{1}{c|}{$\hat{u} = \dfrac{1}{2 \gc} \pdv{\tp}{\xr} z (z-d)$} & \multicolumn{1}{c|}{\cref{ER0:inf:axi}} & \multicolumn{1}{c|}{\cref{ER0:inf:conical}} & \multicolumn{1}{c|}{$u = \dfrac{1}{2\eta_2} \pdv{\tp}{x} \, (z-d) \, z$} & \multicolumn{1}{c|}{\cref{ER0:inf:pland}} \\
    \hline
    $v$ & \multicolumn{1}{|c|}{\cref{uvLowERinf}} & \multicolumn{1}{c|}{$\hat{v} = \dfrac{1}{2 \hc} \pdv{\tp}{\yr} z (z-d)$} & \multicolumn{1}{c|}{\cref{ER0:inf:axi}} & \multicolumn{1}{c|}{\cref{ER0:inf:conical}} & \multicolumn{1}{c|}{$v = \dfrac{1}{2\eta_2} \pdv{\tp}{y} \, (z-d) \, z$} & \multicolumn{1}{c|}{\cref{ER0:inf:pland}} \\
    \hline
    $w$ & \multicolumn{6}{|c|}{$\dfrac{d'}{d^3} (3d - 2z) z^2$} \\
\hline
\end{tabular}
\caption{
A summary of the thin-film Ericksen--Leslie equations \cref{LM1eq0,LM2eq0,LM3eq0,CMeq0,AM1eq0,AM2eq0} in the limit of small Ericksen number ($\ER\ll1$) in terms of the unknown pressure $\tp$ (and the unknown twist angle $\phi$ in the scenario of conical infinite anchoring).
Expressions for the tilt director angle $\theta$, the twist director angle $\phi$, the pressure $\tp$, and the velocities $u$, $v$ and $w$ are stated.
The transformed coordinates $\xr$, $\yr$, $r$ and $\beta$ are defined by \cref{trans1,trans2}.
}
\label{table:LER}
\end{table}

For the degenerate form of conical infinite anchoring called planar degenerate infinite anchoring,
which corresponds to the anchoring conditions \cref{InfDBC} with $\Thetat = 0$, the thin-film Ericksen--Leslie equations \cref{LM1eq0,LM2eq0,LM3eq0,CMeq0,AM1eq0,AM2eq0} can be written in terms of the unknown pressure $\tilde{p}$ as
\begin{align}\label{ER0:inf:pland}
\begin{gathered}
    u = \dfrac{1}{2\eta_1} \left(\bgy\pdv{\tp}{x} -\bgxy \pdv{\tp}{y}\right)z (z-d), \quad
    v = \dfrac{1}{2\eta_1} \left(\bgx\pdv{\tp}{y} -\bgxy \pdv{\tp}{x}\right)z (z-d), \quad
    w = \dfrac{d'}{d^3} \, (3d-2z)\, z^2 \\
    \pdv{x}\bigg( \bgy \pdv{\tp}{x} -\bgxy\pdv{\tp}{y} \bigg) + \pdv{y}\bigg(\bgx \pdv{\tp}{y} -\bgxy\pdv{\tp}{x} \bigg) = \dfrac{12\eta_1 d'}{d^3}, \quad
    \theta \equiv 0, \quad
    \pdv[2]{\phi}{x} + \pdv[2]{\phi}{y} = 0,
\end{gathered}
\end{align}
where $\bgx(\phi)=\gx(0,\phi)=\eta_1 \cos^2 \phi +\sin^2\phi$, $\bgy(\phi)=\gy(0,\phi)=\eta_1 \sin^2 \phi +\cos^2\phi$, and $\bgxy(\phi)=\gxy(0,\phi)=\eta_1 \sin \phi \cos\phi - \sin \phi \cos\phi$. In general, the solution of this set of equations is found by first solving the Laplace equation for $\phi$ subject to a boundary condition on $\partial \Omega$, then substituting the solution for $\phi$ into the differential equation for $\tilde{p}$ and solving for $\tilde{p}$ subject to an appropriate boundary condition on $\partial \Omega$.

\cref{table:LER} summarises the scenarios we have considered in the limit of small Ericksen number $(\ER\ll1)$.

\section{The limit of large Ericksen number ($\ER \gg 1$)}
\label{sec:ERinf}

In the limit of large Ericksen number ($\ER\gg1$), viscous effects are much stronger than elasticity effects and there are two distinct cases to consider, namely when the nematic is a flow-aligning nematic or a non-flow-aligning nematic. These two cases, which we consider in \cref{sec:ERinf:FA,sec:ERinf:NFA}, respectively, arise at leading order in $\ER\gg 1$ from the thin-film angular momentum equations \cref{AM1eq0,AM2eq0}, which are satisfied by either
\begin{equation}
m(\theta) = 0 \quad \text{and} \quad \sin\phi \, \pdv{u}{z} - \cos\phi \, \pdv{v}{z} = 0  \label{scenario1}
\end{equation}
or
\begin{equation}
q(\theta) = 0\quad \text{and} \quad \cos\phi \, \pdv{u}{z} + \sin\phi \, \pdv{v}{z} = 0. \label{scenario2}
\end{equation}

From the definitions of $m(\theta)$ and $q(\theta)$ in \cref{mqfunc} it is clear that, for general values of the viscosity parameters $\gamma_1$ and $\gamma_2$, solutions satisfying \cref{scenario1} or \cref{scenario2} are mutually exclusive. Also, we see from the definition of $m(\theta)$ in \cref{mqfunc} that a solution satisfying \cref{scenario1} is only possible when $-\gamma_2 > \gamma_1$. A material that obeys this condition is known as a flow-aligning nematic \cite{ISBOOK}.
When a solution satisfying \cref{scenario1} is not possible, \ie when $\gamma_1 > -\gamma_2$, the nematic material is known as a non-flow-aligning nematic or a tumbling nematic \cite{ISBOOK}, and so a solution satisfying \cref{scenario2} is then required.
We note that solutions to \cref{scenario1} or \cref{scenario2} do not satisfy the anchoring conditions discussed in \cref{sec:BC}, the anchoring being broken by the flow effects; however, these leading order in $\ER\gg 1$ equations provide the leading-order solutions away from the boundaries of the region and the boundary conditions will be satisfied via appropriate boundary layers \cite{Cousins2020,Crespo2017}, as discussed in the next section.

\subsection{Flow-aligning nematics}
\label{sec:ERinf:FA}

For a flow-aligning nematic, the definition of $m(\theta)$ given in \cref{mqfunc} yields the well-known flow-alignment solution, $\theta \equiv \pm \theta_{\rm L}$, where
\begin{align}
    \thetaL &= \dfrac{1}{2}\cos^{-1}\bigg(-\dfrac{\gamma_1}{\gamma_2}\bigg)
\end{align}
is the Leslie or flow-alignment angle \cite{ISBOOK}.
A stability analysis of the full system \cref{AM1eq,AM2eq,LM1eq,LM2eq,LM3eq} has previously shown that in regions of positive or negative shear rate, the director angle prefers to align at the positive Leslie angle $\thetaL$ or the negative Leslie angle $-\thetaL$, respectively, \cite{ISBOOK}.
As has also been previously demonstrated \cite{Anderson2015,Crespo2017,Cousins2020}, this solution is an outer solution (\ie a solution in the bulk of the cell away from the plates and away from any internal location of director reorientation) and the solution for $\theta$ has boundary layers of thickness $\Oh(\ER^{-1/2}) \ll 1$ near to the plates, in which the director adjusts to satisfy the anchoring conditions at the plates, and an internal layer of thickness $\Oh(\ER^{-1/3}) \ll 1$ near the centre of the cell, which separates the regions of positive and negative shear rate and positive Leslie angle $\thetaL$ and negative Leslie angle $-\thetaL$, respectively. In the present work, we use this outer solution and assume that the internal layer (and therefore the change in the sign of shear rate) is located at $z=d/2$. Hence, at leading-order in $\ER\gg 1$, the solution for $\theta$ is given by
\begin{align}\label{ERinf:thetasol}
    \theta =
    \begin{cases}
    \: +\theta_{\rm L} \quad \text{when} \quad 0 \le z \le d/2, \\
    \: -\theta_{\rm L} \quad \text{when} \quad d/2 < z \le d.
    \end{cases}
\end{align}
The solution for $\phi$ satisfies the second equation of \cref{scenario1}. Therefore, using the approach detailed in \cref{sec:ER0:pat}, at leading order in $\ER\gg 1$, the thin-film Ericksen--Leslie equations \cref{LM1eq0,LM2eq0,LM3eq0,CMeq0,AM1eq0,AM2eq0} can now be written in terms of the unknown pressure $\tp$ as
\begin{align}\label{ERinf:FA}
\begin{gathered}
    u = \dfrac{1}{2\etaL} \pdv{\tp}{x} z (z-d), \quad
    v = \dfrac{1}{2\etaL} \pdv{\tp}{y} z (z-d), \quad
    w = \dfrac{d'}{d^3} \, (3d-2z)\, z^2, \\
    \pdv[2]{\tp}{x} + \pdv[2]{\tp}{y} = \dfrac{12 \etaL d'}{d^3}, \quad
    \theta =\pm \thetaL, \quad
    \tan \phi = \dfrac{v}{u},
\end{gathered}
\end{align}
where $\etaL=g(\thetaL)=g(-\thetaL)$ is the local effective viscosity of a flow-aligned nematic, which can be defined in terms of the Miesowicz viscosities as
\begin{align}
    \etaL &= \dfrac{\eta_1}{2} \left(1- \dfrac{\gamma_1}{\gamma_2} \right) + \dfrac{\eta_2}{2} \left(1+ \dfrac{\gamma_1}{\gamma_2} \right) +\dfrac{\eta_{12}}{4} \left(1- \dfrac{\gamma_1}{\gamma_2} \right)\left(1+ \dfrac{\gamma_1}{\gamma_2} \right).
\end{align}
Therefore, we find that, at leading order in $\ER\gg 1$, the flow of a flow-aligning nematic is identical to the flow of an isotropic fluid with effective viscosity $\etaL$, with the behaviour of the director determined by the behaviour of the flow; that is, the director lies in the plane that contains the flow direction and the direction of maximum shear stress, and aligns at the Leslie angle from the flow direction.

\subsection{Non-flow-aligning nematics}
\label{sec:ERinf:NFA}

\begin{table}[tp]
\centering
\setlength\tabcolsep{0.1cm}
\renewcommand{\arraystretch}{1.75}
\begin{tabular}{|c|c|c|}
    \hline
    & \multicolumn{2}{c|}{Large Ericksen number $\ER \gg 1$} \\ \cline{2-3}
    & \: Flow-aligning nematic (\cref{sec:ERinf:FA}) \: & \: Non-flow-aligning nematic (\cref{sec:ERinf:NFA}) \: \\ \cline{2-3}
    \: $\theta$ \: & $\pm \thetaL$ & $0$ \\[0.15cm] \hline
    $\phi$ & $\tan^{-1} \bigg( \dfrac{v}{u}\bigg)$ & $\tan^{-1} \bigg( -\dfrac{u}{v}\bigg)$ \\[0.15cm] \hline
    $\tp$ & $\pdv[2]{\tp}{x} + \pdv[2]{\tp}{y} = \dfrac{12 \etaL d'}{d^3}$ & $\pdv[2]{\tp}{x} + \pdv[2]{\tp}{y} = \dfrac{12 d'}{d^3}$ \\[0.15cm] \hline
    $u$ & $\dfrac{1}{2\etaL} \pdv{\tp}{x} \, (z-d) \, z$ & $\dfrac{1}{2}\pdv{\tp}{x} \, (z-d) \, z$\\[0.15cm] \hline
    $v$ & $\dfrac{1}{2\etaL} \pdv{\tp}{y} \, (z -d) \,z$ & $\dfrac{1}{2}\pdv{\tp}{y} \,(z -d) \,z$ \\[0.15cm] \hline
    $w$ & \multicolumn{2}{c|}{$\dfrac{d'}{d^3} (3d - 2z)z^2$} \\[0.15cm] \hline
\end{tabular}
\caption{
A summary of the thin-film Ericksen--Leslie equations \cref{LM1eq0,LM2eq0,LM3eq0,CMeq0,AM1eq0,AM2eq0} in the limit of large Ericksen number in terms of the unknown pressure $\tp$. Expressions for the tilt director angle $\theta$, twist director angle $\phi$, the equation governing the pressure $\tp$, and the velocities $u$, $v$ and $w$ are stated.
}
\label{table:HER}
\end{table}

For a non-flow-aligning nematic, the two solutions to \cref{scenario2} are
\begin{equation}
    \tan\phi=-\frac{u}{v} \quad \mbox{with} \quad \theta=0 \quad \mbox{or} \quad \theta=\frac{\pi}{2}.
\end{equation}
Of these two possibilities, the stable solution is $\theta=0$ with $\tan\phi=-u/v$, known as the log-rolling solution \cite{ISBOOK}, which has been studied in detail theoretically by Alonso \etal \cite{Alonso2003} and experimentally by Romo-Uribe and Windle \cite{Romo-Uribe1996}. Therefore, at leading order in $\ER\gg 1$, the thin-film Ericksen--Leslie equations \cref{LM1eq0,LM2eq0,LM3eq0,CMeq0,AM1eq0,AM2eq0} can be written in terms of the unknown pressure $\tp$ as
\begin{align}\label{ERinf:NFA}
\begin{gathered}
    u = \dfrac{1}{2} \pdv{\tp}{x} z (z-d), \quad
    v = \dfrac{1}{2} \pdv{\tp}{y} z (z-d), \quad
    w = \dfrac{d'}{d^3} \, (3d-2z)\, z^2, \\
    \pdv[2]{\tp}{x} + \pdv[2]{\tp}{y} = \dfrac{12 d'}{d^3}, \quad
    \theta = 0, \quad
    \tan \phi = -\dfrac{u}{v},
\end{gathered}
\end{align}
and we find that, at leading order in $\ER\gg 1$, the flow of a non-flow-aligning nematic is identical to the flow of an isotropic fluid with unit effective viscosity, which corresponds to a dimensional viscosity $\eta_3$, with the behaviour of the director determined by the behaviour of the flow; that is, it aligns perpendicular to both the flow direction and the direction of maximum shear stress.

\cref{table:HER} summarises the scenarios we have considered in the limit of large Ericksen number ($\ER \gg 1$).

\section{Application to the One Drop Filling method}
\label{sec:app}

\begin{figure}[tp]
\begin{center}
\includegraphics[width=0.7\linewidth]{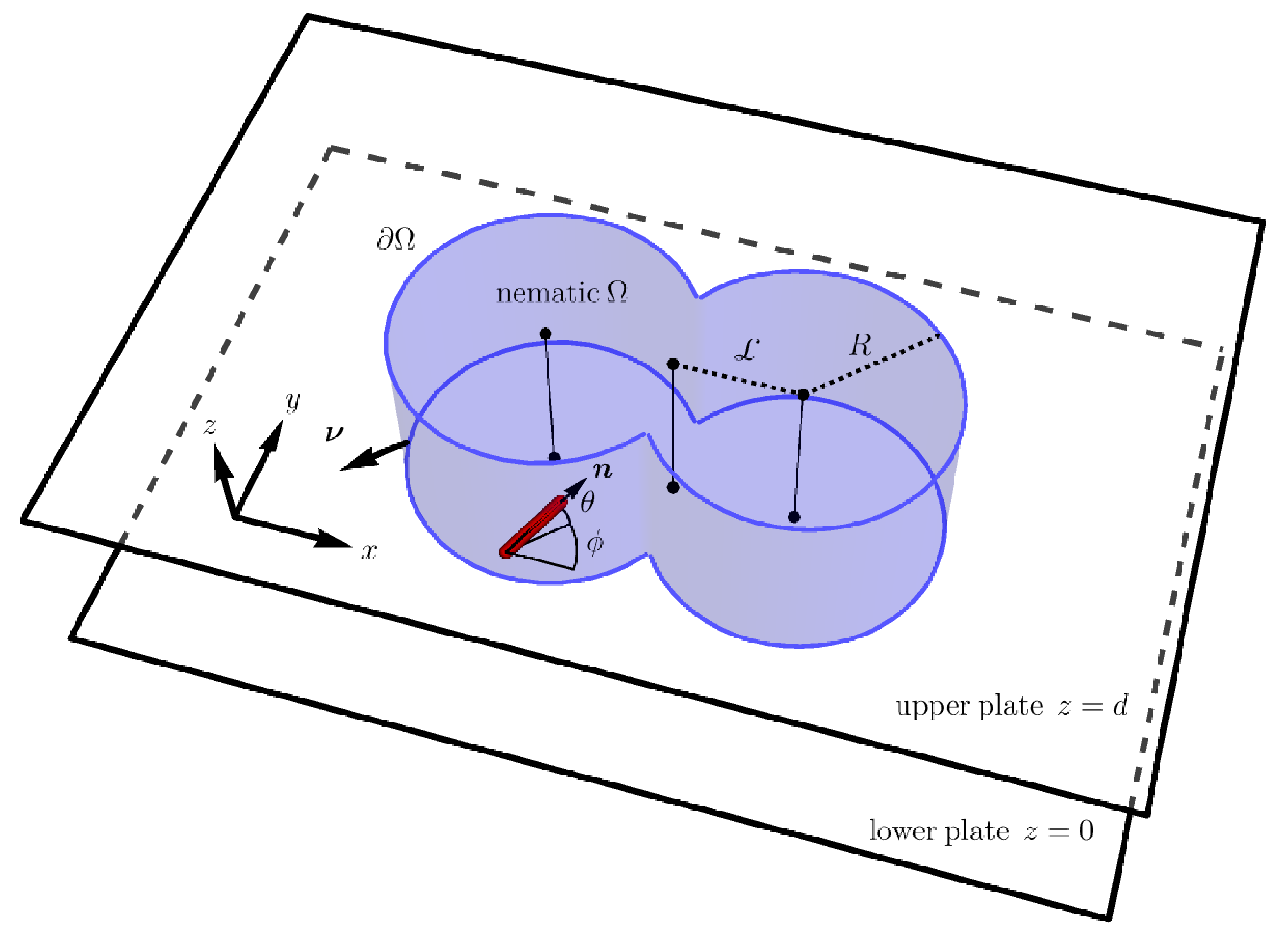}
\end{center}
\caption{
A Hele-Shaw cell showing a perspective view of a region of nematic $\Omega$ with free surface $\partial \Omega$ with outward unit normal $\bm{\nu}$ defined by two overlapping cylindrical droplets (in light blue) bounded between solid parallel plates at $z=0$ and $z=d$.
The Cartesian coordinates $x$, $y$ and $z$, the tilt director angle $\theta$, the twist director angle $\phi$, the equal cylindrical radii $R$, the fixed half separation of the cylindrical droplets $\mathcal{L}$, the fixed axes of the cylindrical droplets $(-\mathcal{L},0,z)$ and $(\mathcal{L},0,z)$, and the fixed axis on which the centre of mass lies $(0,0,z)$ are also shown.
}
\label{figDrop}
\end{figure}

As an example of the insight that can be gained by using the present theoretical approach, we calculate the flow that occurs during the squeezing of two coalescing nematic regions sandwiched between two parallel plates, as depicted in \cref{figDrop}. This situation is a simple model for the squeezing stage of the ODF method, as discussed in \cref{sec:intro}, in which regions of nematic are forcibly coalesced as an upper plate, located at $d(t)$, is moved towards a fixed lower plate, located at $z=0$, In the present situation we assume that the decreasing gap between the plates is given, in dimensional form, by
\begin{align}
    d = d_0 - \Sp t \quad \text{for} \quad 0\le t \le t_{\rm f}, \label{H:ODF}
\end{align}
where $d_0$ is the initial thickness of the cell, $\Sp$ is the speed at which the upper plate moves toward the lower plate, and $t_{\rm f} = (d_0 - d_{\rm f})/\Sp$ is the time at which the required final thickness of the cell $d_{\rm f}$ is achieved.

In the model presented in this section, we neglect any transient initial inertial effects at the start of the squeezing process and consider the two limiting cases of small and large Ericksen numbers. We also assume that the quasi-static evolution of the free surface $\partial \Omega = \partial \Omega(t)$ is prescribed using the solution of a conservation-of-volume model of two overlapping cylindrical nematic droplets $\Omega = \Omega(t)$. The model therefore neglects any effects that surface tension, elasticity, anchoring, and contact line dynamics might have on the evolution of the free surface.
Specifically, we assume that the nematic region is the union of two cylindrical regions of nematic, having equal radii $R = R(t)$ and fixed axes $(-\mathcal{L},0,z)$ and $(\mathcal{L},0,z)$, with a combined constant volume $V$, and outward unit normal $\bm{\nu} = \bm{\nu}(x,y,t)$, as shown in \cref{figDrop}. The centre of mass of the nematic region lies on the fixed axis $(0,0,z)$, and there are cusps in the free surface formed where the overlapping cylinders meet at $(0,-c,z)$ and $(0,c,z)$, where $c = \sqrt{R(t)^2 - \mathcal{L}^2}$.
The volume of the nematic region is then given by \cite{Cousinsthesis}
\begin{align}
    \dfrac{V}{d} &= \Bigg[\pi - \cos^{-1} \bigg( \dfrac{\mathcal{L}}{R}\bigg) \Bigg] R^2 +\mathcal{L} \sqrt{R(t)^2 - \mathcal{L}^2}, \label{R:eq}
\end{align}
and, due to the conservation of volume, the evolution of the radius $R$, which determines the shape of the free surface $\partial\Omega$ is given implicitly by \cref{R:eq}.
The boundary condition for $u$ and $v$ on $\partial\Omega$ is then given by the kinematic condition
\begin{align}
    \bm{u} \cdot \bm{\nu} = R' \quad \text{on} \quad \partial \Omega, \label{odf:bc}
\end{align}
where $R' = \dd R /\dd t$ is the speed at which the free surface expands, which may be obtained via implicit differentiation of \cref{R:eq} with respect to $t$.

Free surfaces formed between a nematic and air (or vacuum) have been found to exhibit a variety of types of anchoring \cite{SONINBOOK}. However, since homeotropic anchoring is the most commonly reported anchoring at free surfaces \cite{SONINBOOK,Cousins2023}, and because this type of anchoring at a free surface is exhibited by a key component of modern nematic mixtures used in the ODF method, namely 4'-pentyl-4-biphenylcarbonitrile (5CB) \cite{E7Mixture}, we take the anchoring condition on $\partial\Omega$ to be homeotropic infinite anchoring, such that
\begin{align} \label{ProblemLER}
    \bm{\nu} \cdot \bm{n} = 1 \quad \text{on} \quad \partial \Omega.
\end{align}
The anchoring conditions on the plates are assumed to be planar degenerate infinite anchoring (as discussed in \cref{sec:ER0:conP}).
Note that, in the limit of large Ericksen number, which we will discuss shortly, the behaviour of the director field is determined by the behaviour of the flow in the bulk of the nematic region, and the anchoring conditions on the free surface and plates are not required.

An alternative approach to specifying the boundary conditions at the free surface $\partial\Omega$ could be to consider the balances of stress and torque as well as the kinematic condition on $\partial \Omega$.
For example, considering the balance of normal stress on an isotropic free surface leads to the well-known isotropic Young--Laplace equation, which allows the effect of surface tension to influence the shape of the free surface \cite{Crowdy2001,Shelley1997}.
The boundary conditions on a nematic free surface can be considerably more complex than in the isotropic case and involve the combined effects of surface tension, elasticity, anchoring, and contact line dynamics \cite{Cousins2022,Rey2007}.
However, Cousins \etal \cite{Cousins2018} previously used the relatively simple model described above to make qualitative comparisons between theoretical predictions for the speed at which the free surface expands and experimental photographs of ODF mura, demonstrating that the timescale of coalescence due to surface tension effects is much longer than the timescale of the ODF squeezing process. This previous work did not calculate the flow or director field within the nematic region but, as we shall see shortly, this simple model of the free surface also leads to solutions for the director field that compare well to experimental results.

Before considering the limits of small and large Ericksen numbers, we introduce the appropriate non-dimensionalisation for the flow of nematic during the squeezing stage of the ODF method.
In particular, we take the characteristic lengthscale in the $z$-direction to be the initial separation of the plates $\Hscale = d_0$, the characteristic lengthscale in the $xy$-plane to be half the separation of the droplets, which is also the initial radius of the cylindrical regions, $L = \mathcal{L} = R_0 = \sqrt{V/(\pi d_0)}$, and the characteristic velocity scale to be the velocity scale for a flow driven by squeezing a circular cylindrical volume $V$ of nematic between parallel plates $U = \Hvscale\sqrt{V/(4\pi\Hscale^3)}$, where the characteristic plate speed $\Hvscale$ is given by \cref{H:ODF} as $\Hvscale=\Sp$.
The aspect ratio $\delta$, the Ericksen number $\ER$, and the reduced Reynolds $\delta \RE$ can then be written in terms of $d_0$, $\Sp$, $V$ and the nematic material parameters, namely
\begin{align} \label{ODFgroups}
\begin{gathered}
    \delta =\sqrt{\dfrac{\pi d_0^3}{V}}, \quad
    \ER = \dfrac{\mu \Sp}{K} \sqrt{\dfrac{V}{4\pi d_0}} \quad \text{and} \quad \delta \RE = \dfrac{\rho \Sp d_0}{2 \mu}.
\end{gathered}
\end{align}
Using the typical parameter values for the ODF method listed in \cref{table:ODF} and the parameters values of the nematic 5CB, namely $\mu=\eta_3=0.0326\,$Pa$\,$s$^{-1}$ and $K = 6.1\,$p$\,$N \cite{PHYSICALPROPERTIES}, with \cref{ODFgroups} yields $\delta = 0.017$, $\ER = 2.2 \times 10^3$ and $\delta \Re = 0.017$.
These values are consistent with those for the ODF method listed in \cref{table:examples}, as well as with the assumptions $\delta \ll 1$, $\delta \RE \ll 1$ and $\ER \gg 1$, and therefore the limit of large Ericksen number is likely to be the most appropriate. However, for completeness and to compare the two different limiting behaviours, we will also consider the behaviour in the limit of small Ericksen number.

\begin{table}[tp]
\begin{center}
\centering
\begin{tabular}{|c|c|c|c|c|c|}
    \hline
    Parameter & $d_0$ & $V$ & $\Sp$ & $R_0$ &  $d_{\rm f}$ \\
    \hline
    \ Typical value \qquad & \ $74.0 \, \mu$m \qquad & \ $4.5\,\mu$l \qquad & \ $1.0\,$mm$\,$s$^{-1}$ \qquad & \ $4.4\,$mm \qquad & \ $5.0 \, \mu$m \qquad \\
    \hline
\end{tabular}
\caption{
Typical parameter values for the ODF method, specifically the initial thickness of the cell $d_0$, the volume of the region consisting of two overlapping cylindrical nematic regions $V$, the speed at which the upper plate moves downward towards the lower plate $\Sp$, the initial radius of each droplet $R_0$, and the final thickness of the cell $d_{\rm f}$ \cite{Cousinsthesis,Cousins2018}.
}
\label{table:ODF}
\end{center}
\end{table}

\subsection{The limit of small Ericksen number}
\label{sec:app:LER}

\begin{figure}[tp]
\begin{center}
\begin{tabular}{ccc}
\hspace{-0.6cm}
\includegraphics[width=0.36\linewidth]{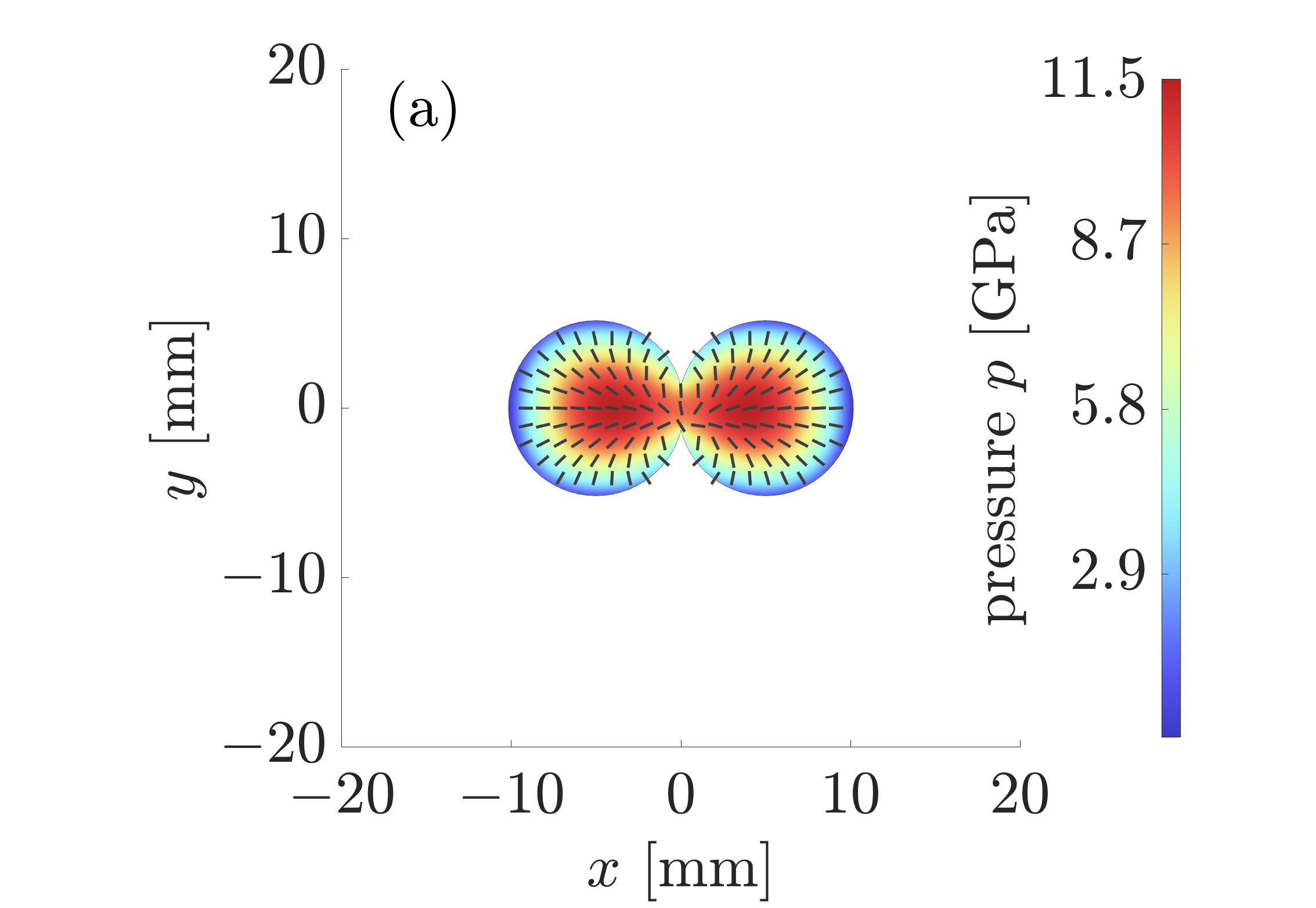} &
\hspace{-0.5cm}
\includegraphics[width=0.36\linewidth]{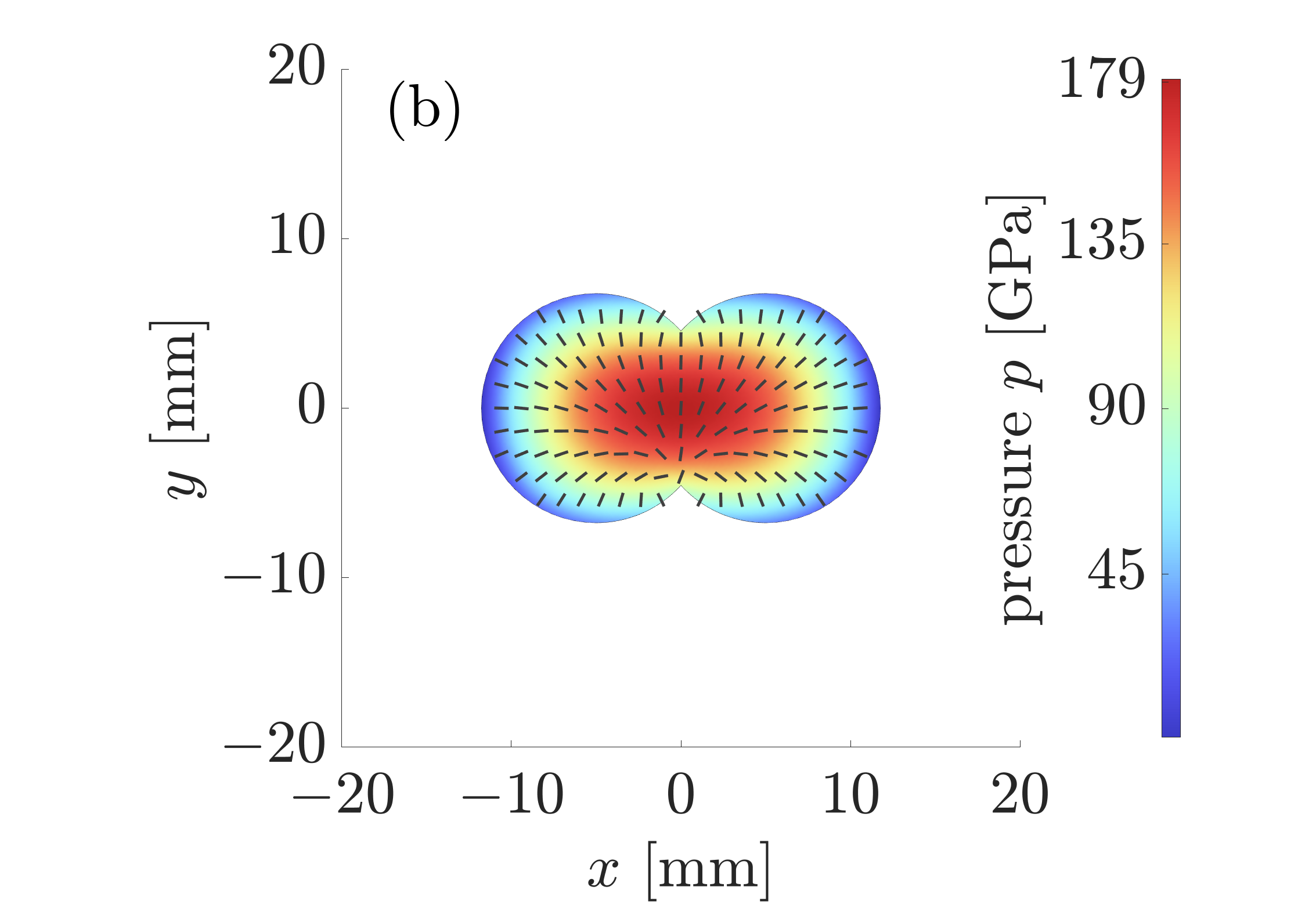} &
\hspace{-0.5cm}
\includegraphics[width=0.36\linewidth]{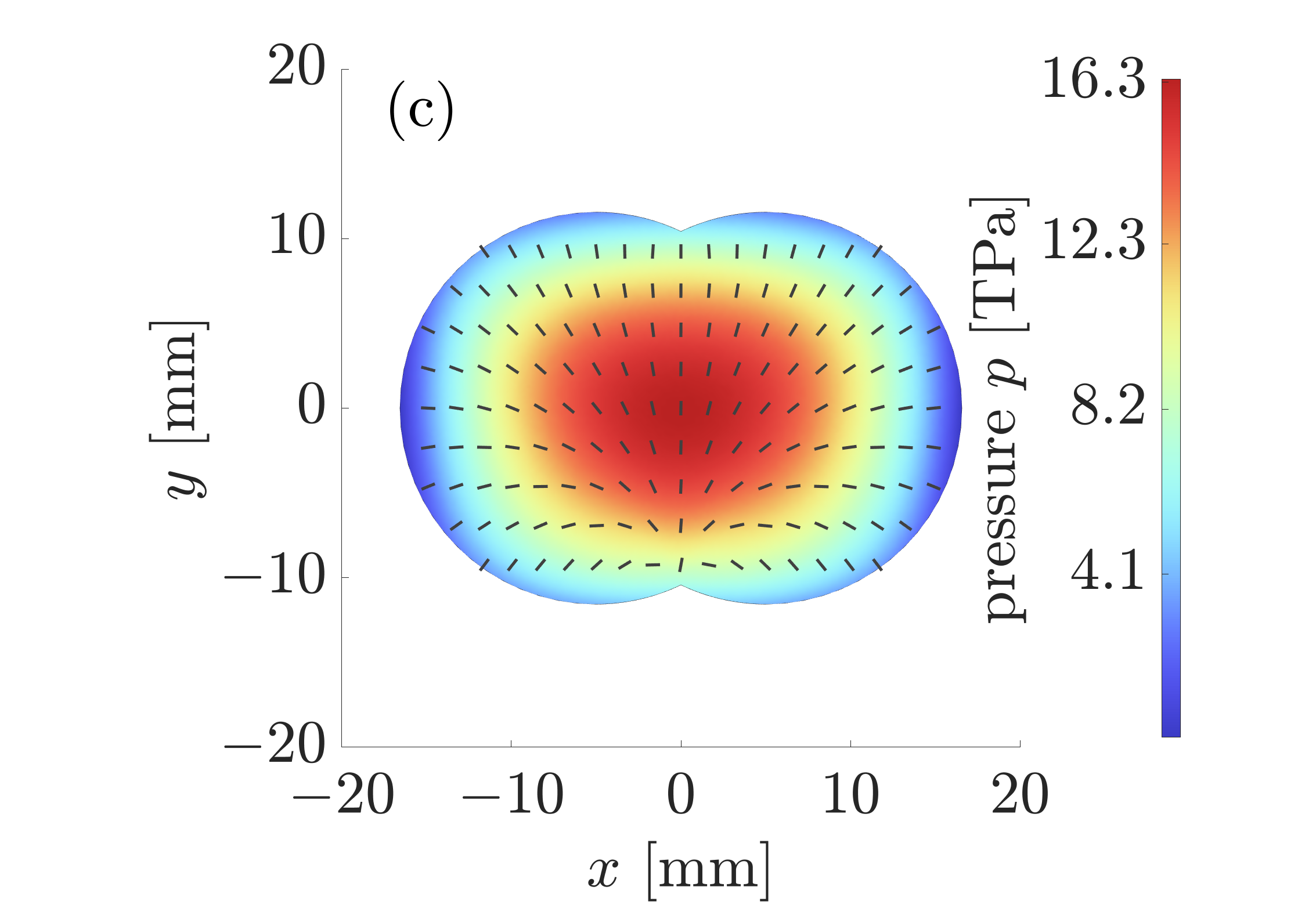} \\
\hspace{-0.6cm}
\includegraphics[width=0.36\linewidth]{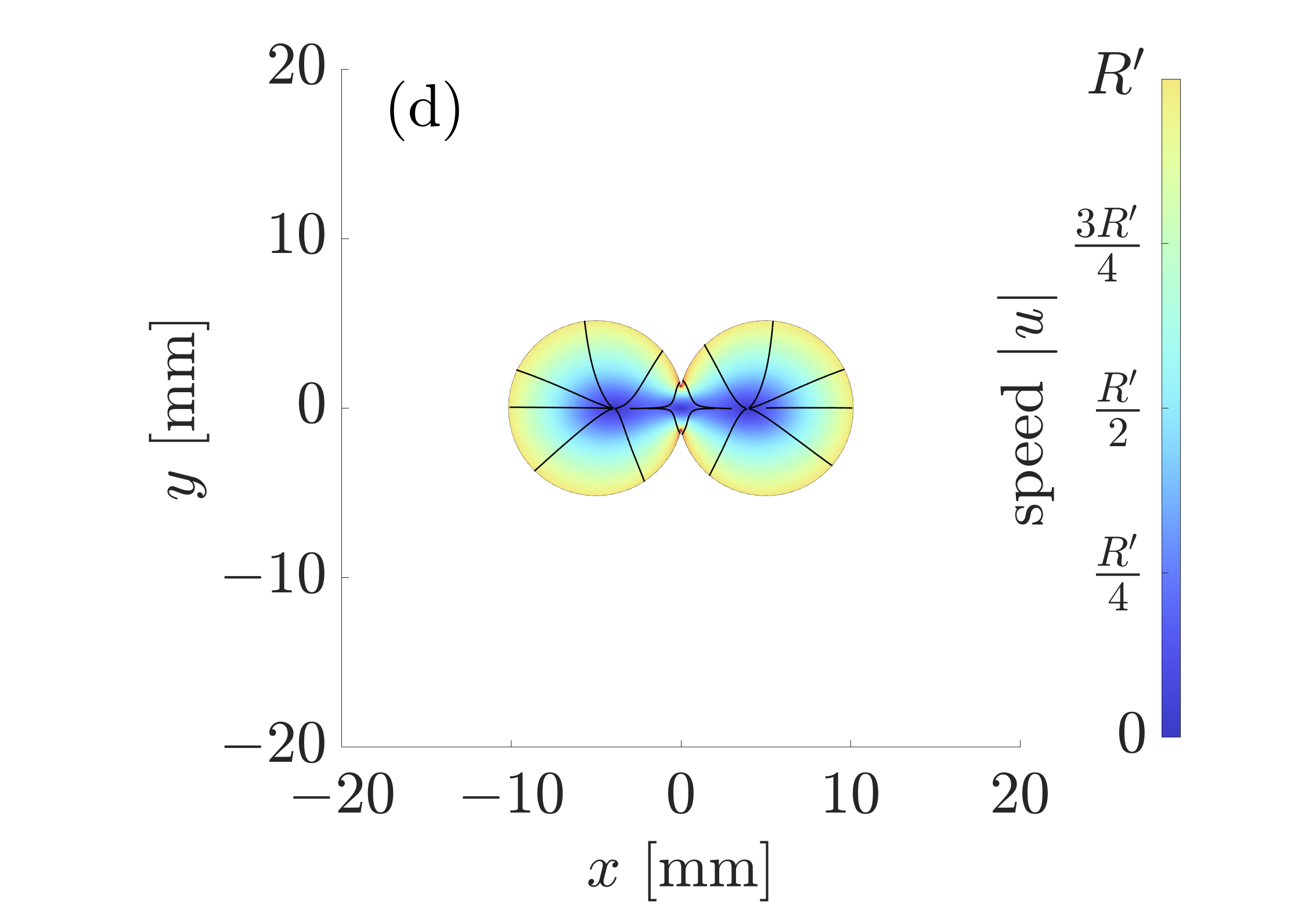} &
\hspace{-0.5cm}
\includegraphics[width=0.36\linewidth]{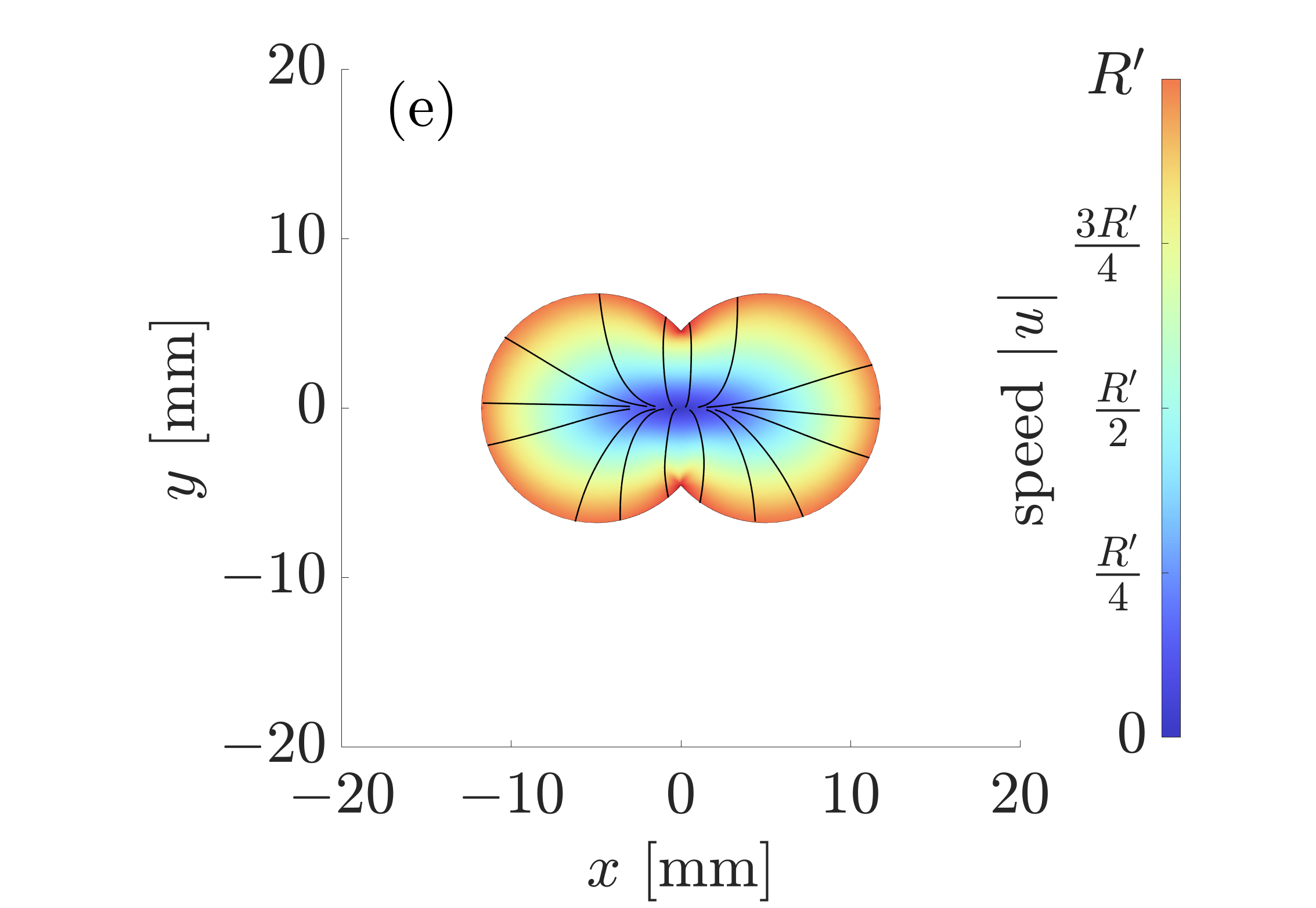} &
\hspace{-0.5cm}
\includegraphics[width=0.36\linewidth]{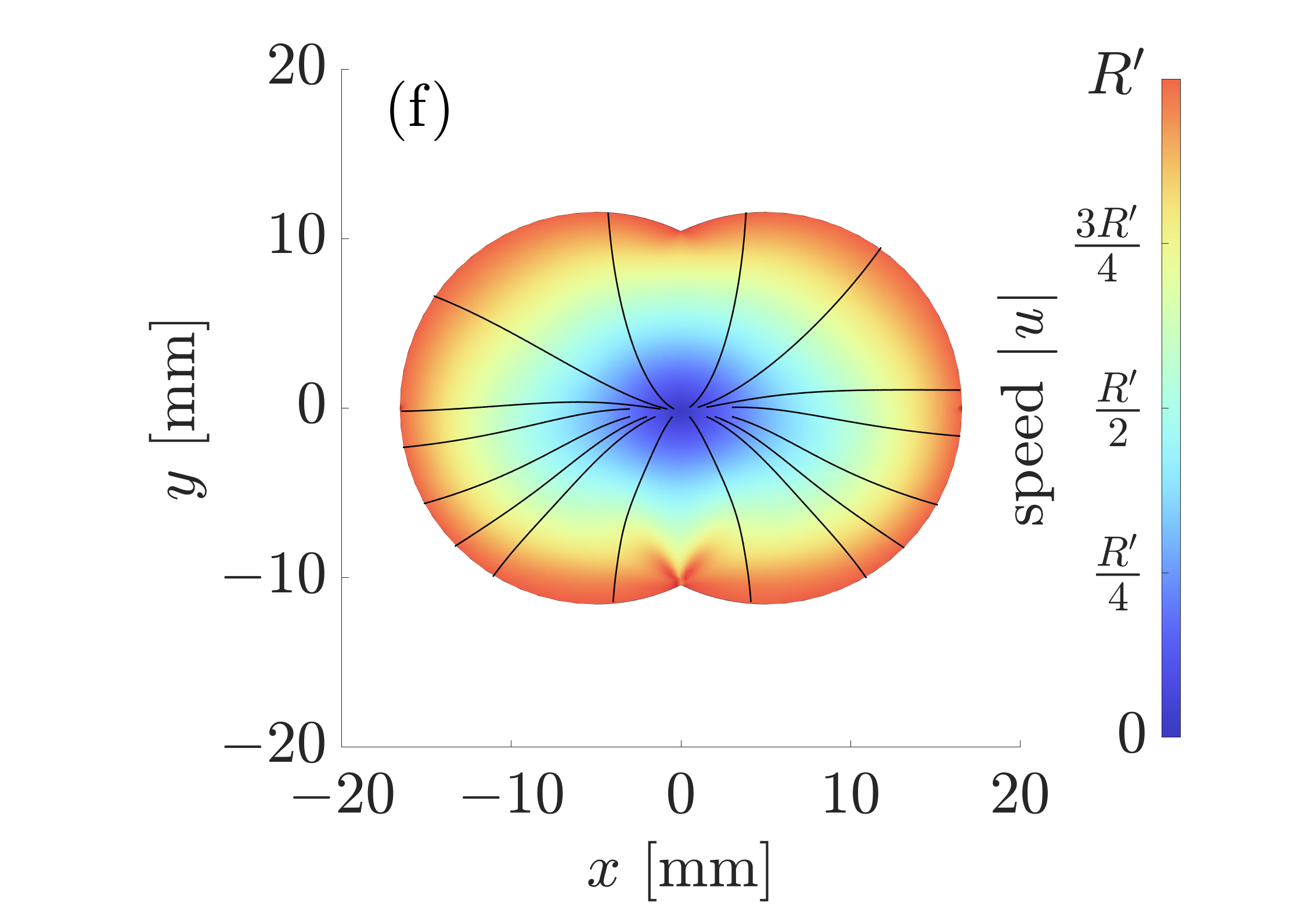} \\
\hspace{-0.6cm}
\includegraphics[width=0.36\linewidth]{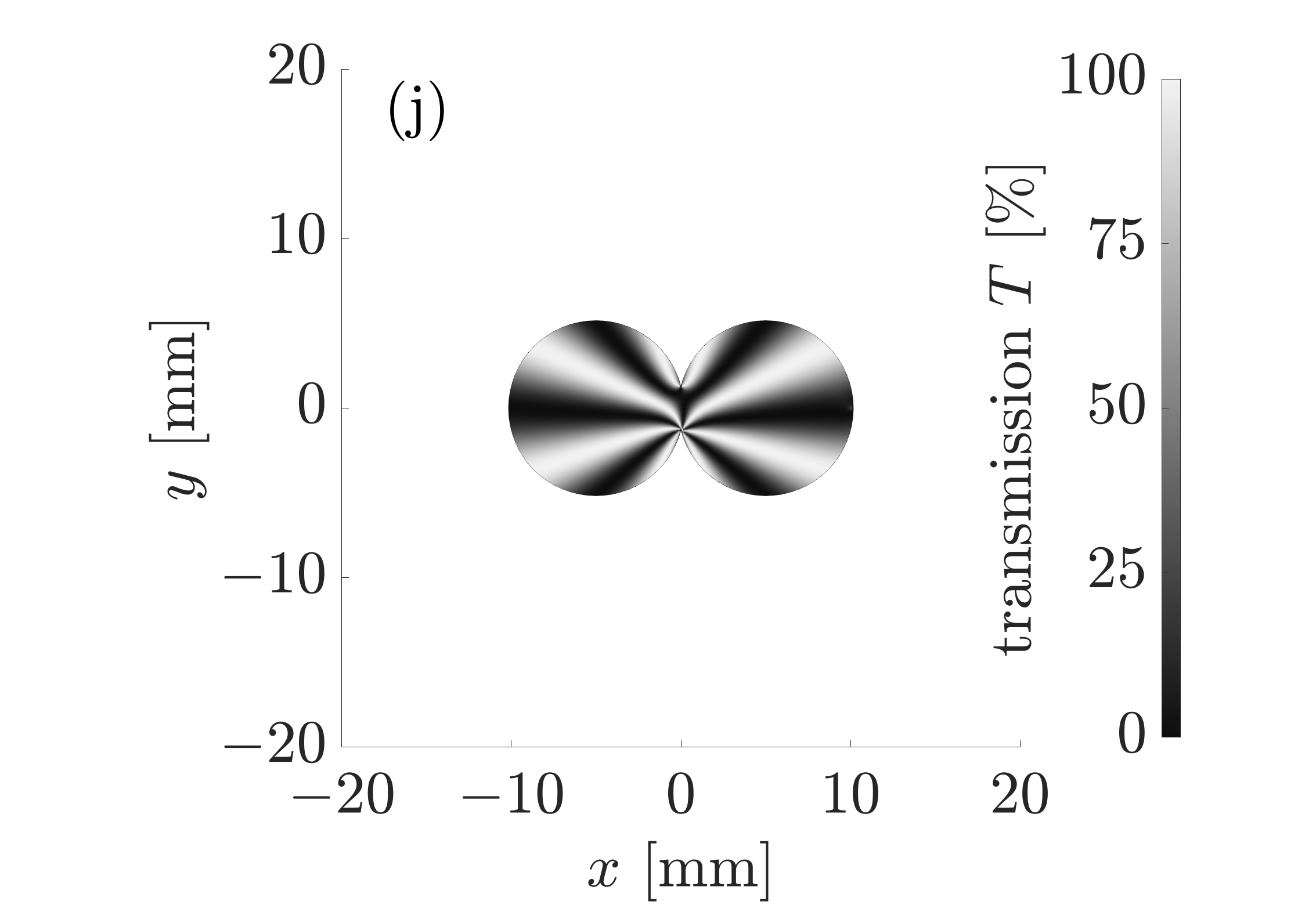} &
\hspace{-0.5cm}
\includegraphics[width=0.36\linewidth]{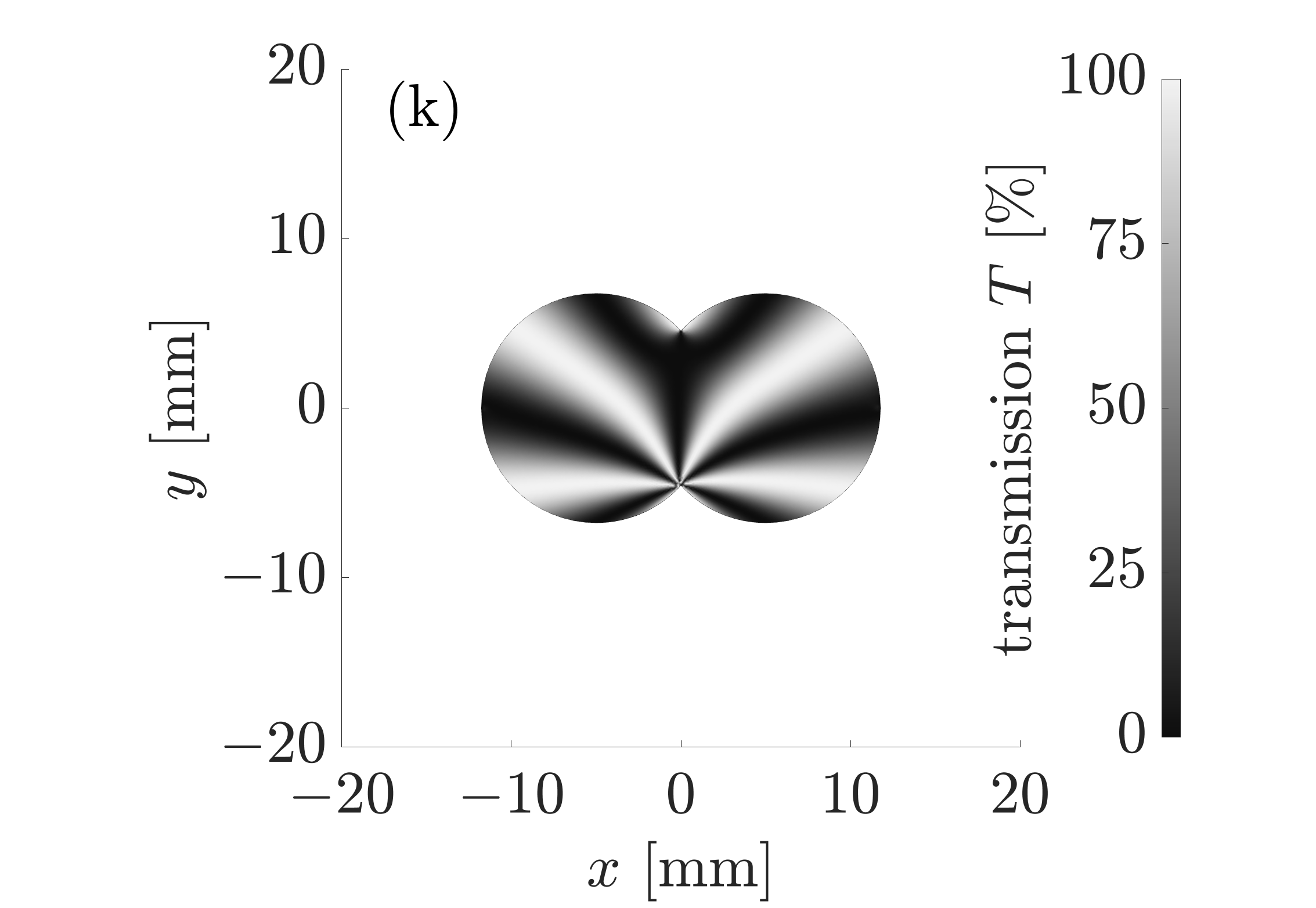} &
\hspace{-0.5cm}
\includegraphics[width=0.36\linewidth]{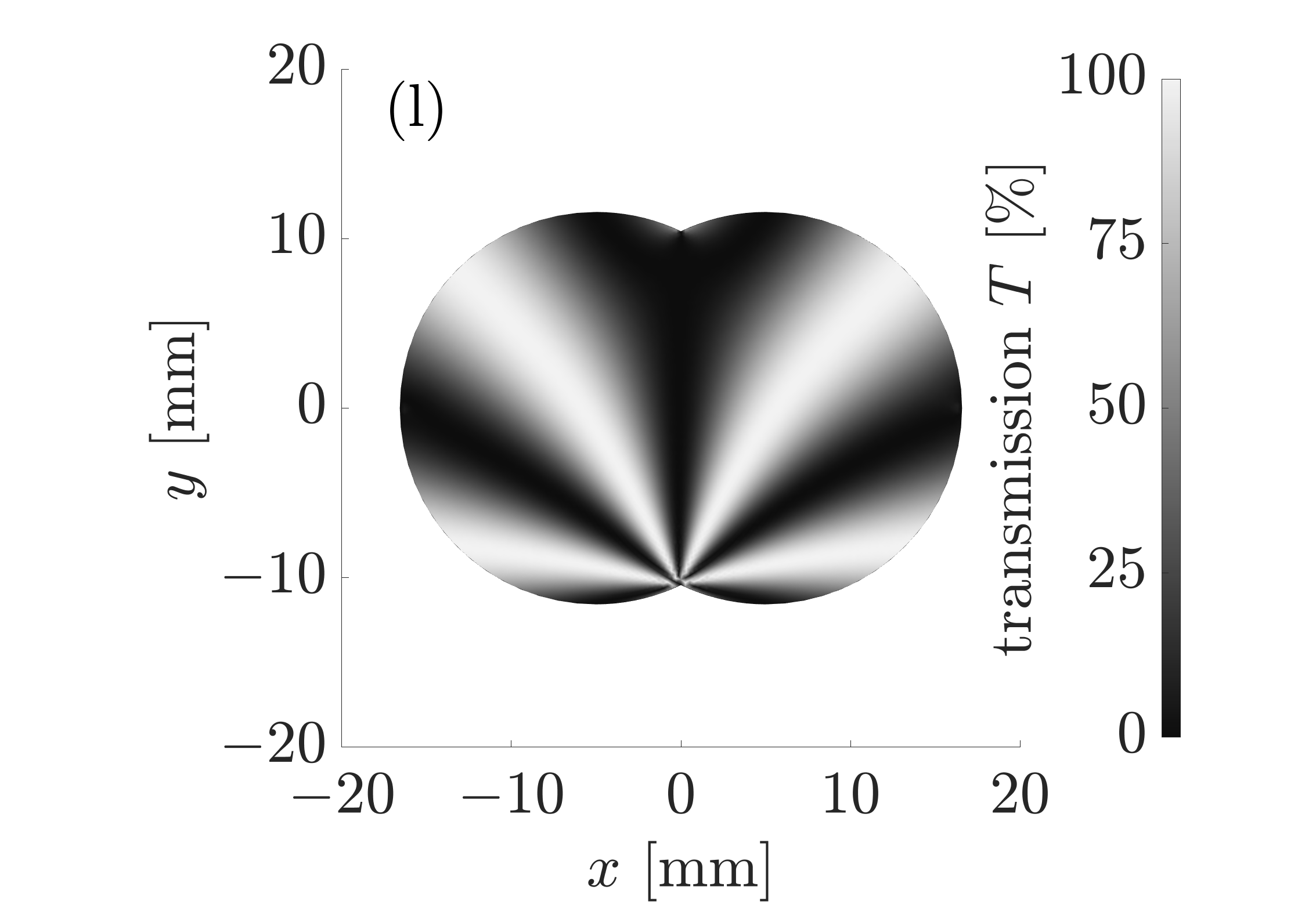}
\end{tabular}
\caption{
Top row - the pressure $\tp$ (coloured background) and the director field $\n$ (black rods); Middle row - flow speed $|\vel|$ (coloured background) and streamlines (solid black lines); Bottom row - the optical transmission $T$ (greyscale): in the limit of small Ericksen number ($\ER \ll 1$) for two coalescing droplets at: $t = 0.02 \,$s when $d = 54\, \mu {\rm m}$ and $R' = 0.05\,$m$\,$s$^{-1}$ (first column, (a), (d) and (j)); $t=0.04 \,$s when $d = 34 \, \mu {\rm m}$, and $R' = 0.12\,$m$\,$s$^{-1}$ (second column, (b), (e) and (h)); and $t = 0.06 \,$s when $d = 14 \, \mu {\rm m}$ and $R' = 0.49\,$m$\,$s$^{-1}$ (third column, (c), (f) and (i)).
}
\label{fig:LER}
\end{center}
\end{figure}

In the limit of small Ericksen number ($\ER\ll1$), and with planar degenerate anchoring on the plates, the appropriate equations are provided in the final column of \cref{table:LER}. We proceed by first solving Laplace's equation for the twist angle $\phi$ given by \cref{ER0:inf:pland} subject to the anchoring condition \cref{ProblemLER}, and then solving the equation for the pressure $\tp$ given by \cref{ER0:inf:pland} subject to the kinematic condition \cref{odf:bc}.
We note that \cref{phiLaplace,ProblemLER} can be solved analytically using a conformal mapping from two intersecting circles to the half-plane; however, this approach leads to integrals that must be evaluated numerically, and so there is little advantage of pursuing this approach over a purely numerical approach.
We therefore use COMSOL Multiphysics \cite{COMSOLcode} to numerically solve \cref{phiLaplace,ProblemLER} for the twist director angle $\phi$, and subsequently numerically solve equation \cref{ER0:inf:pland} for the pressure $\tp$.
Once the solutions for $\phi$ and $\tp$ are determined, the expressions for $u$ and $v$ in \cref{ER0:inf:pland} determine the velocity.
From \cref{ER0:inf:pland} we see that a non-uniform solution for the director field will lead to an anisotropic patterned viscosity via the effective viscosity functions $\bgx$, $\bgy$ and $\bgxy$, which themselves depend on the nondimensional viscosity $\eta_1$ (equal to the ratio of dimensional viscosities $\eta_1/\eta_3$).
The flow is driven by the squeezing together of the plates but guided by the patterned viscosity.
Note that, in the special case that the dimensional viscosities $\eta_1$ and $\eta_3$ are equal, the director does not affect the flow.

\cref{fig:LER} shows the numerically calculated solutions for pressure $\tp$ and the director field $\n$ (top row), as well as the speed $|\vel|$ and streamlines (middle row) at leading order in the limit of small Ericksen number ($\ER \ll 1$) for the parameters listed in \cref{table:ODF} at three different times: the first column, (a), (d) and (g), for $t= 0.02 \,$s when $d = 54\, \mu {\rm m}$ and $R' = 0.05\,$m$\,$s$^{-1}$; the second column, (b), (e) and (h), for $t=0.04 \,$s when $d = 34 \, \mu {\rm m}$ and $R' = 0.12\,$m$\,$s$^{-1}$; and the third column, (c), (f) and (i), for $t= 0.06 \,$s when $d= 14\, \mu {\rm m}$ and $R' = 0.49\,$m$\,$s$^{-1}$.
Note that, since the director field is fixed by the infinite anchoring on the free surface, there is a discontinuity in the director field at the cusp located at $(0,-c)$. The symmetry of the system means that there is another solution with a defect at the cusp at $(0,c)$, which is a reflection of the solution in \cref{fig:LER} in the line through the centres of the cylinders. There are also higher energy solutions that satisfy \cref{phiLaplace,ProblemLER},
including a solution with defects at both cusps, but these solutions are unlikely to occur in practice due to their higher energy.
Asymmetry in the director field solution leads to asymmetry in the effective viscosity, and then to asymmetry in the pressure $\tp$ and the velocity, shown by the streamlines in \cref{fig:LER}.

Finally, we calculate an approximation to the (relative) optical transmission, denoted $T=T(x,y)$, through the droplets between crossed polarisers aligned with the $x$-axis and the $y$-axis, which is measured relative to the transmission of light when the component of the director $\n$ in the $xy$-plane is aligned with one of the polarisers \cite{HANDBOOK_LCoptics}.
Specifically, the approximation to the optical transmission is $T = \sin^2 2 \phi$ \cite{HANDBOOK_LCoptics}. With this measure of transmission, $100\%$ optical transmission occurs when the twist angle $\phi$ is $\pi/4$ or $3\pi/4$, \ie when the component of $\n$ in the $xy$-plane is $\pi/4$ from both polarisers, and $0\%$ optical transmission occurs when the twist angle is $0$ or $\pi/2$, \ie when the component of $\n$ in the $xy$-plane is aligned with one of the polarisers.
\cref{fig:LER} (bottom row) shows the optical transmission $T$ in the limit of small $\ER$, and provides a visualisation of the director field that can be readily be compared to the optical transmission observed in experiments.

\subsection{The limit of large Ericksen number}
\label{sec:app:HER}

\begin{figure}[tp]
\begin{center}
\begin{tabular}{ccc}
\hspace{-0.6cm}
\includegraphics[width=0.36\linewidth]{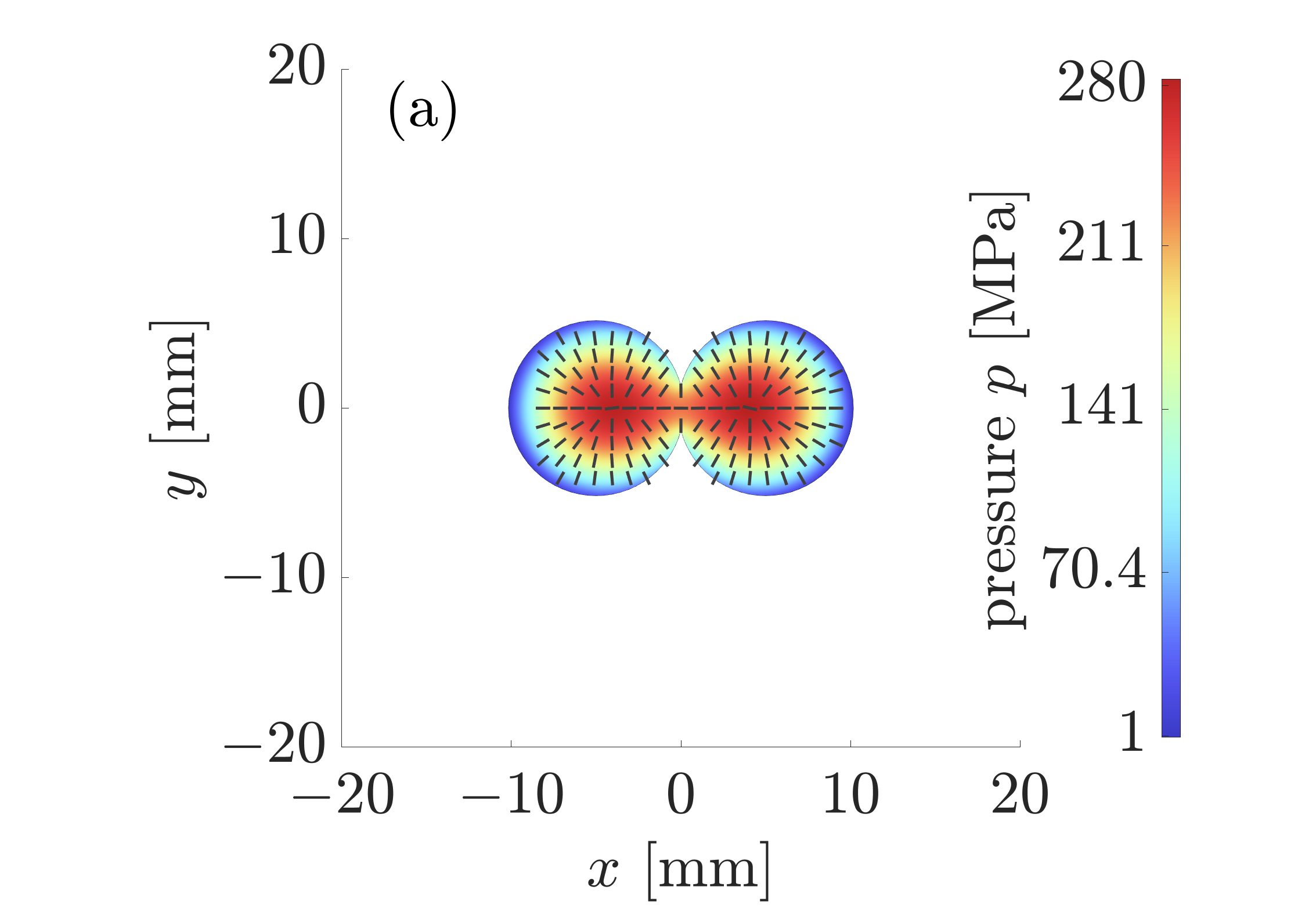} &
\hspace{-0.5cm}
\includegraphics[width=0.36\linewidth]{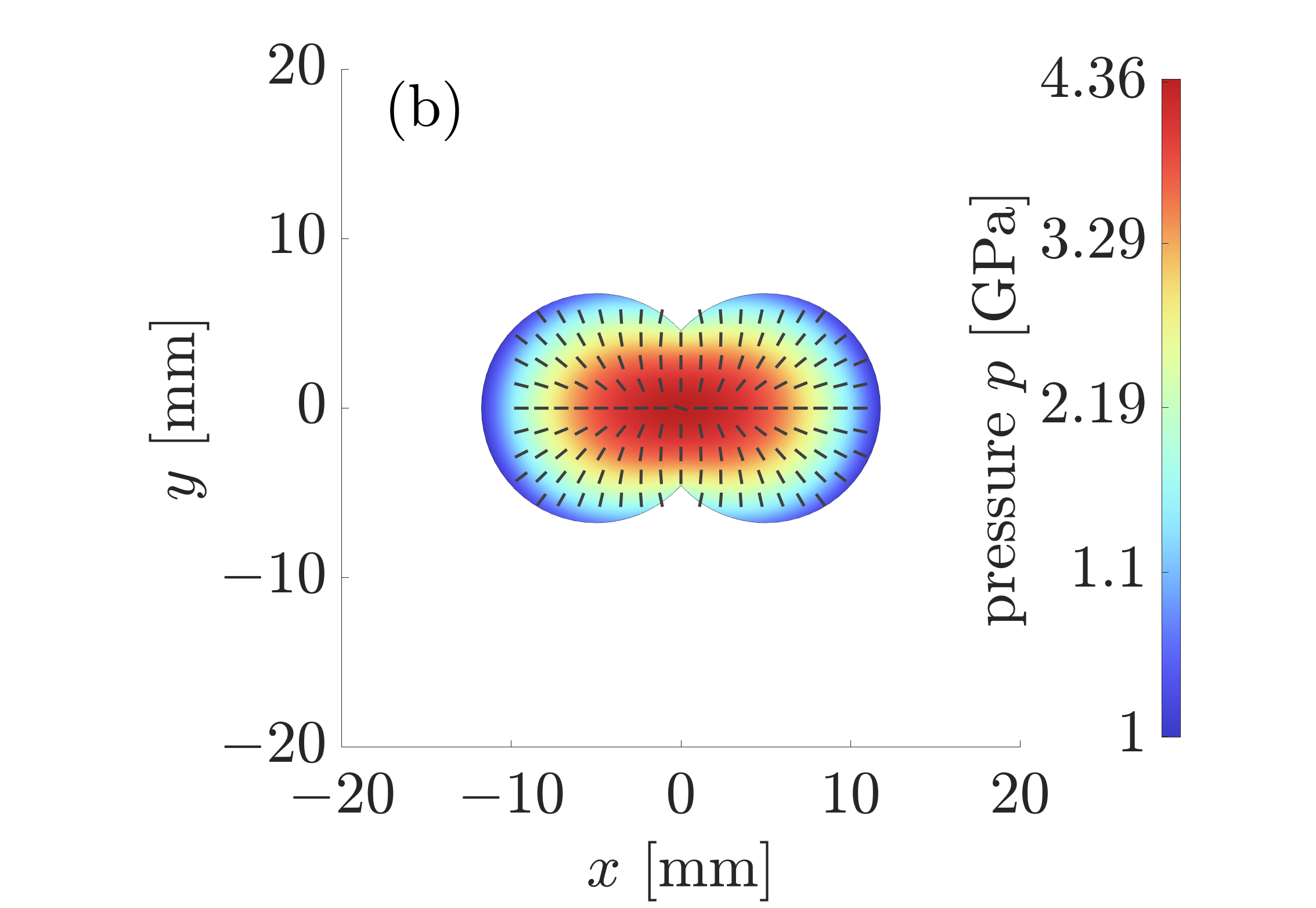} &
\hspace{-0.5cm}
\includegraphics[width=0.36\linewidth]{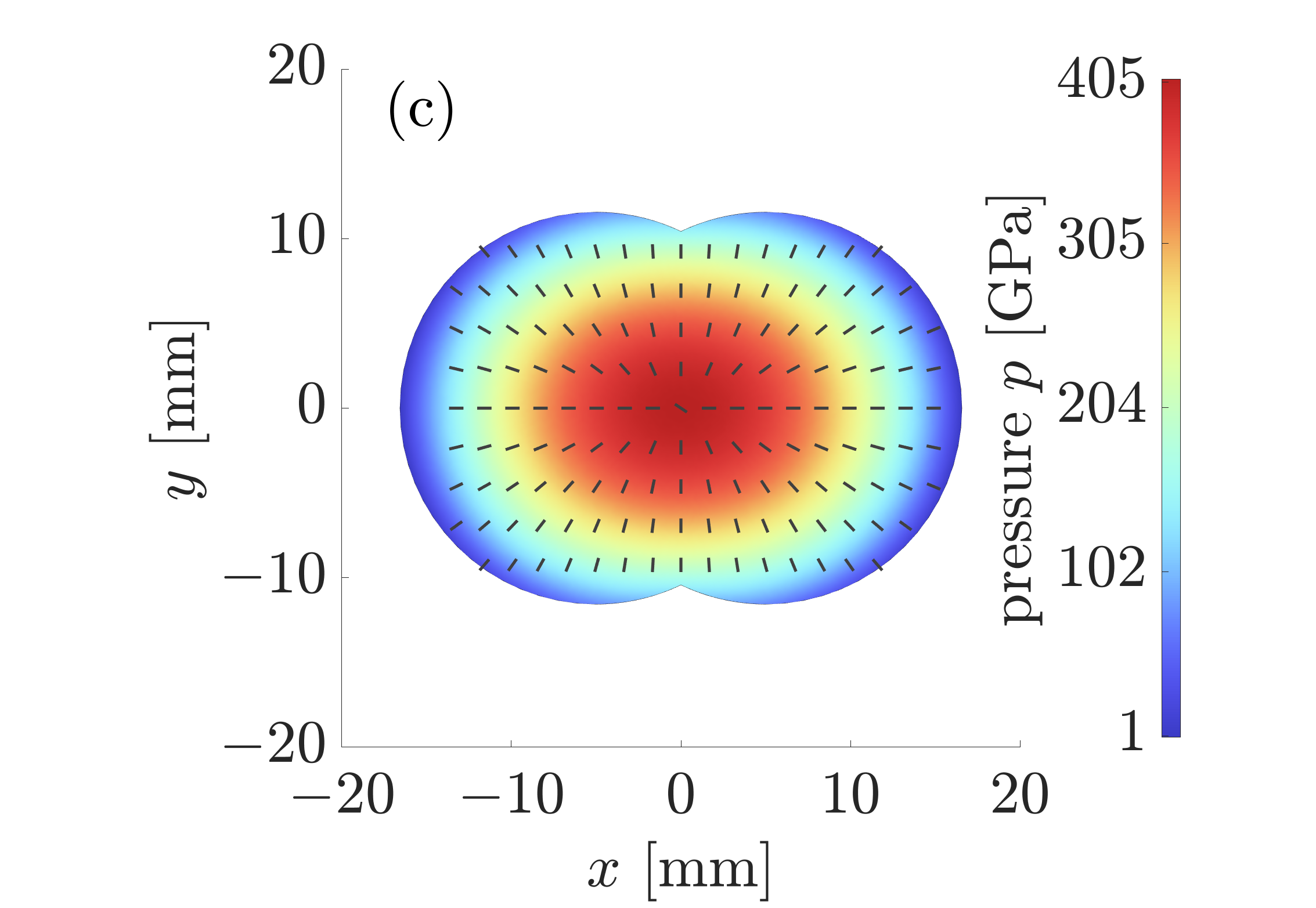} \\
\hspace{-0.6cm}
\includegraphics[width=0.36\linewidth]{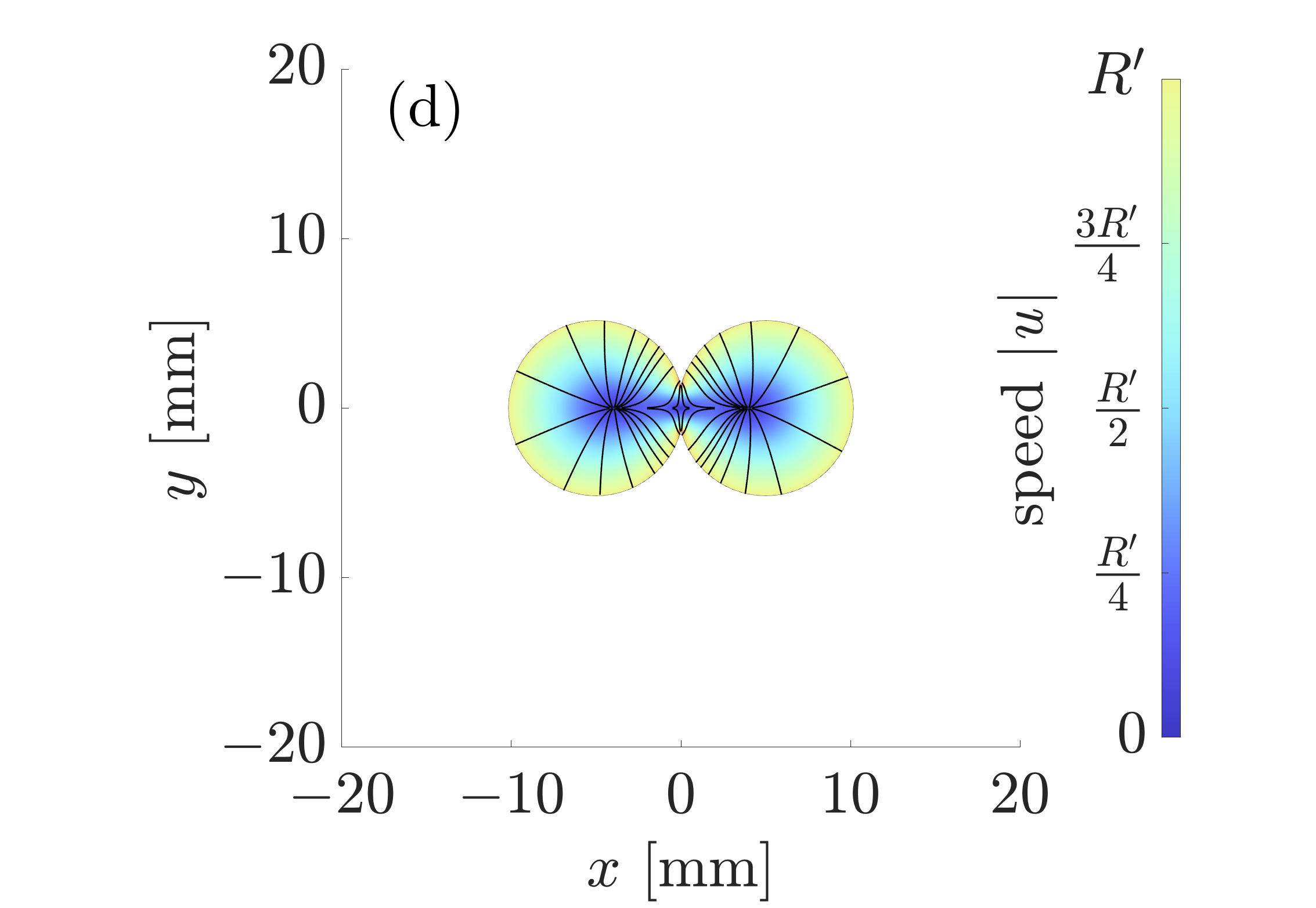} &
\hspace{-0.5cm}
\includegraphics[width=0.36\linewidth]{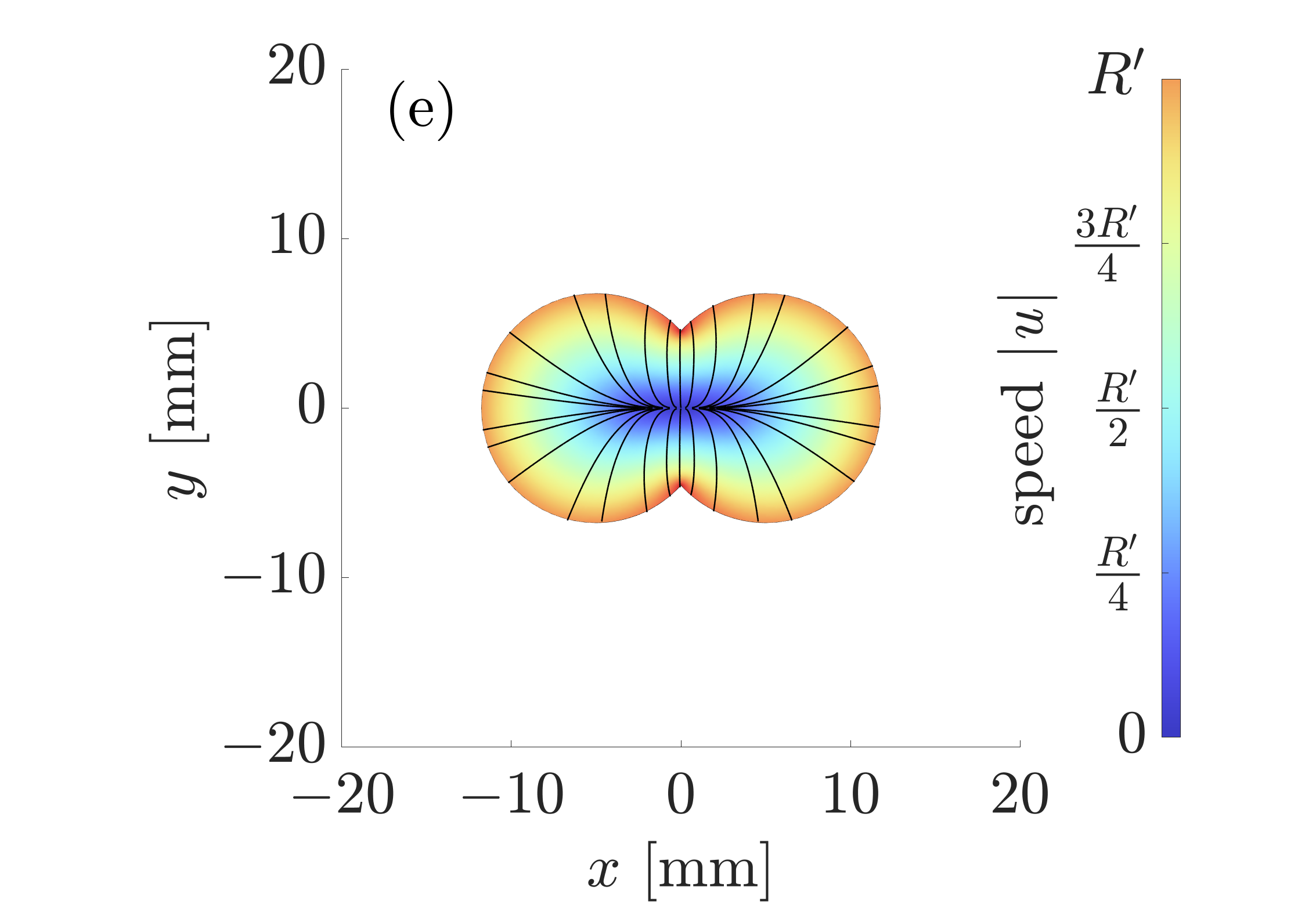} &
\hspace{-0.5cm}
\includegraphics[width=0.36\linewidth]{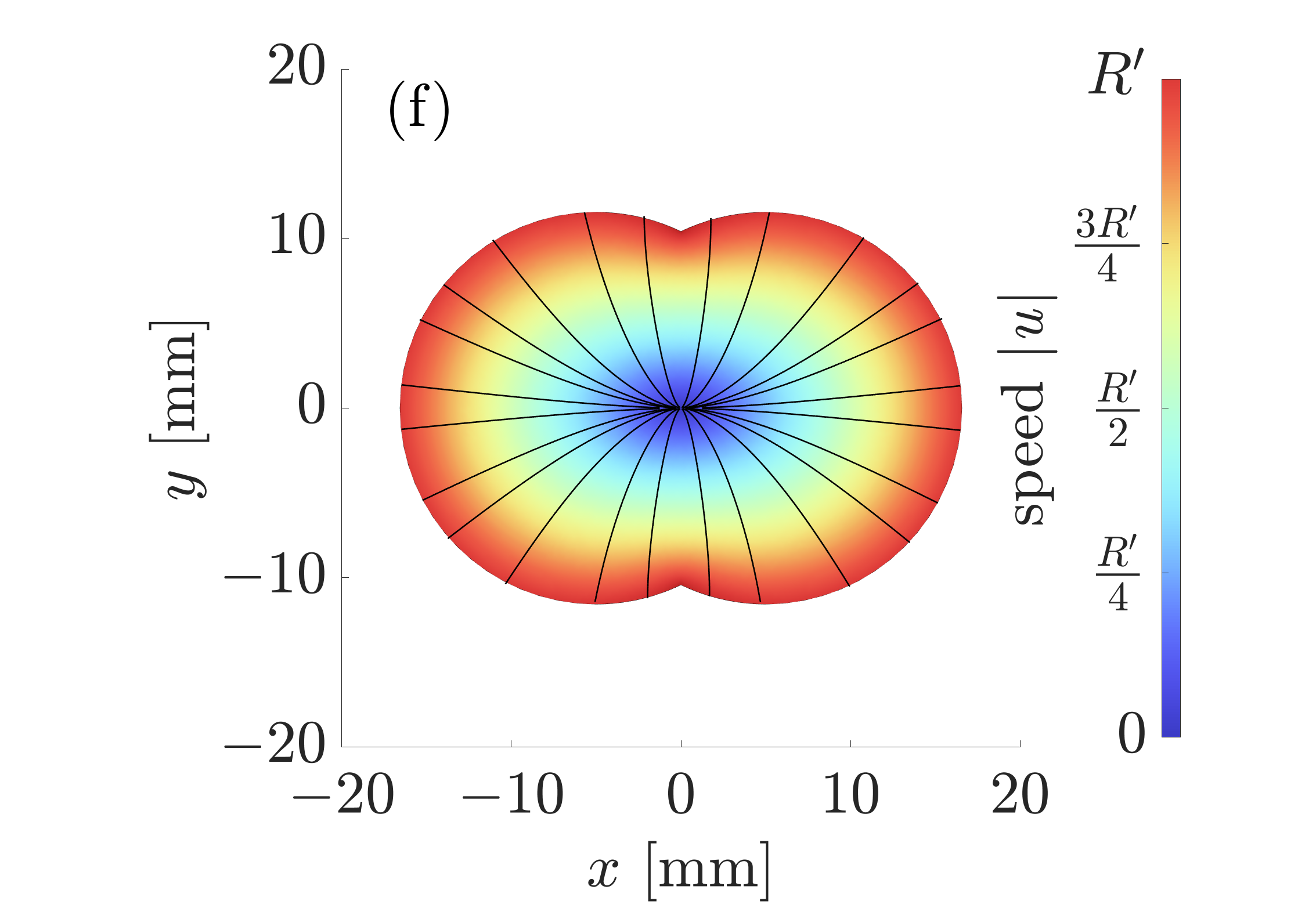} \\
\hspace{-0.6cm}
\includegraphics[width=0.36\linewidth]{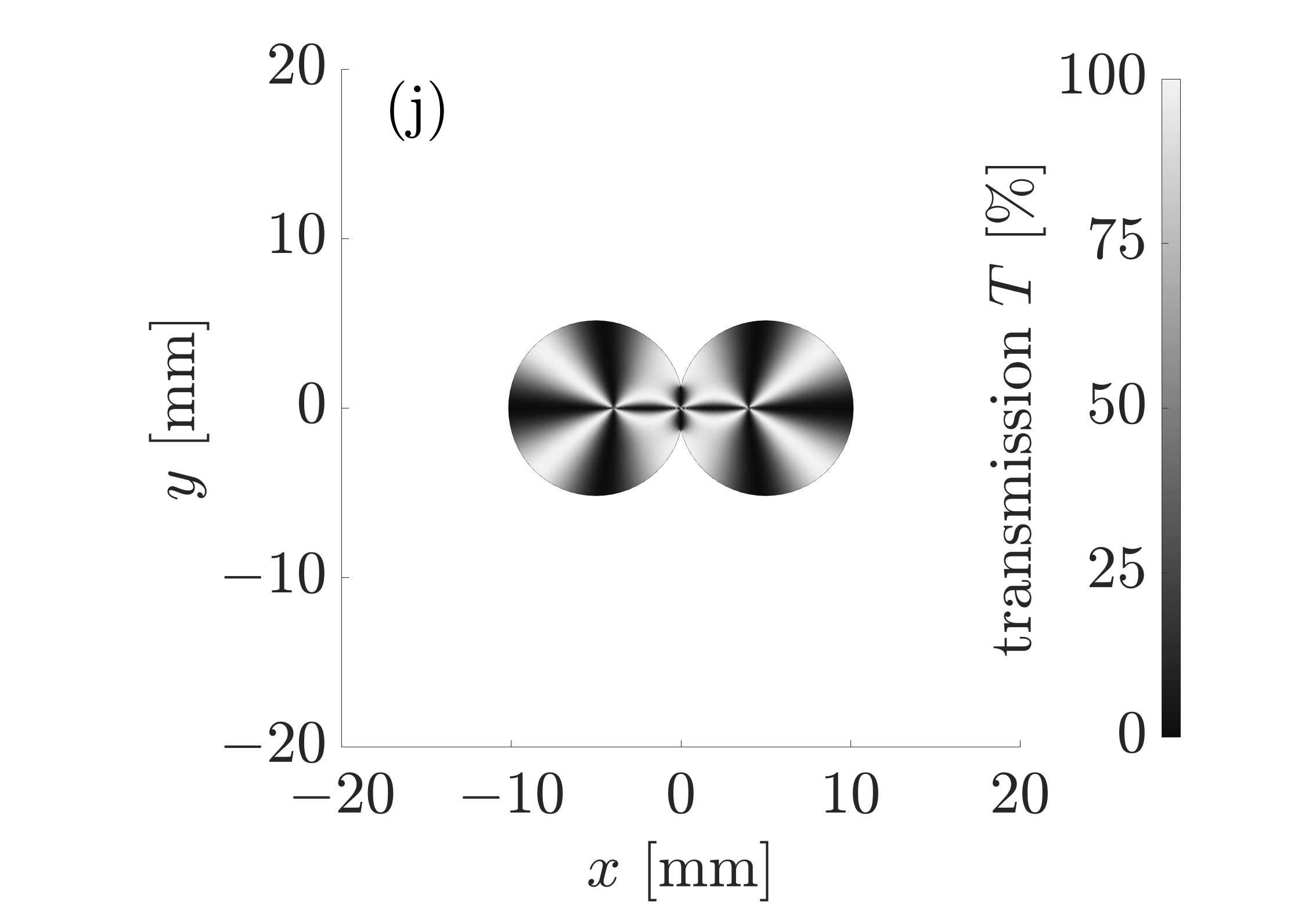} &
\hspace{-0.5cm}
\includegraphics[width=0.36\linewidth]{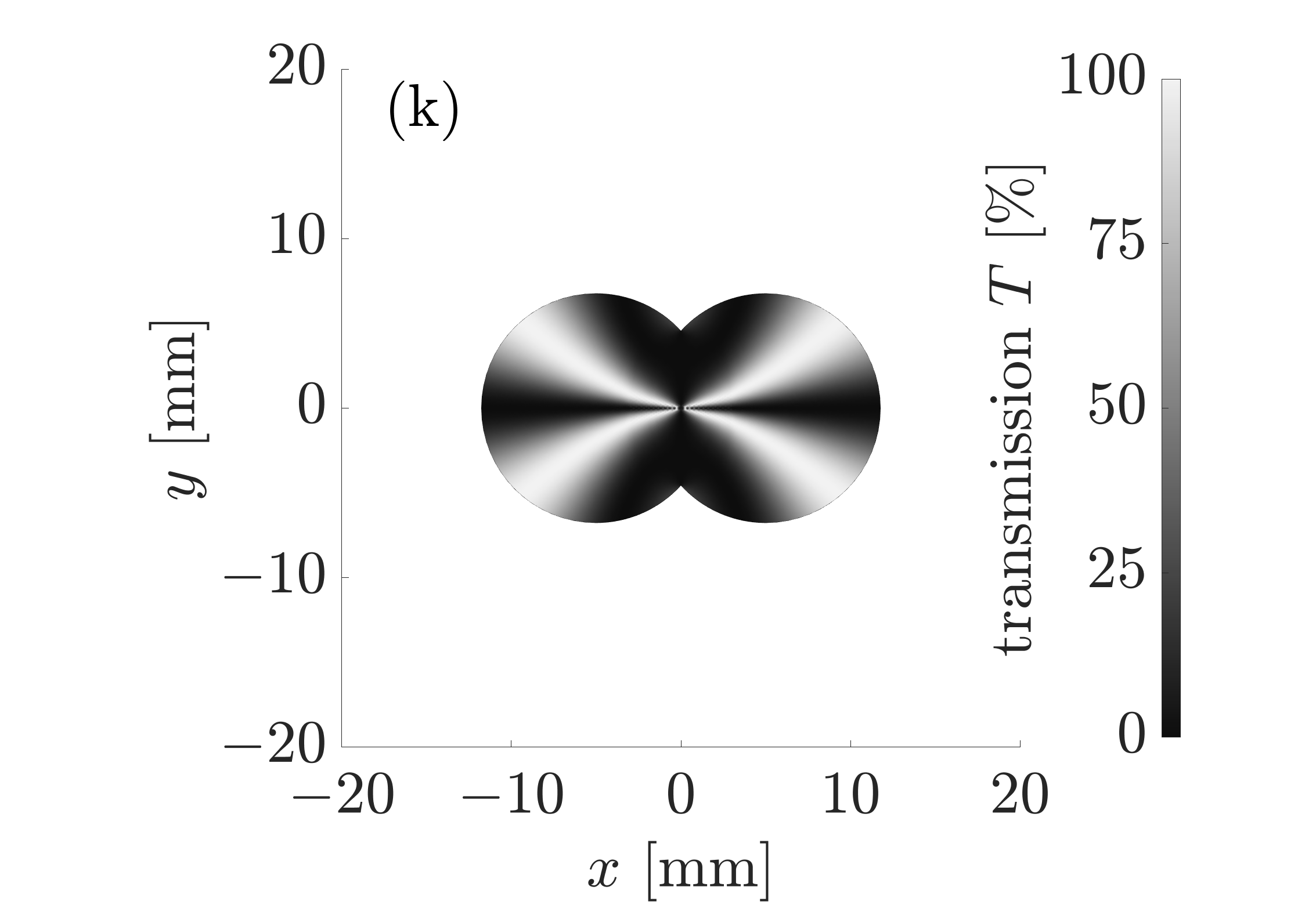} &
\hspace{-0.5cm}
\includegraphics[width=0.36\linewidth]{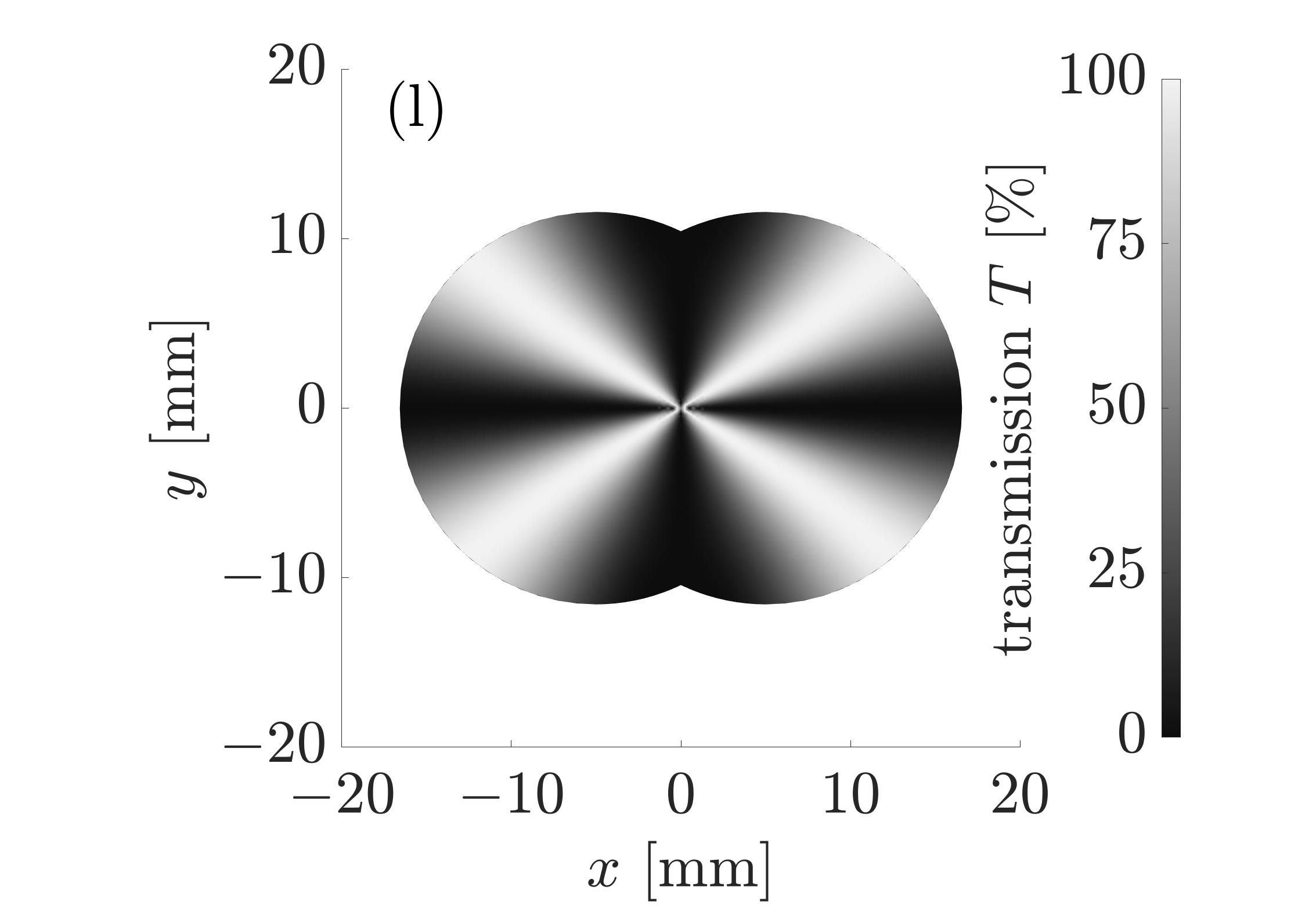}
\end{tabular}
\caption{
Top row - the pressure $\tp$ (coloured background) and the director field $\n$ (black rods); Middle row - flow speed $|\vel|$ (coloured background) and streamlines (solid black lines); Bottom row - the optical transmission $T$ (greyscale): in the limit of large Ericksen number ($\ER \gg 1$) for two coalescing droplets at: $t = 0.02 \,$s when $d = 54\, \mu {\rm m}$ and $R' = 0.05\,$m$\,$s$^{-1}$ (first column, (a), (d) and (j)); $t=0.04 \,$s when $d = 34 \, \mu {\rm m}$, and $R' = 0.12\,$m$\,$s$^{-1}$ (second column, (b), (e) and (h)); and $t = 0.06 \,$s when $d = 14 \, \mu {\rm m}$ and $R' = 0.49\,$m$\,$s$^{-1}$ (third column, (c), (f) and (i)).
}
\label{fig:HER}
\end{center}
\end{figure}

For the limit of large Ericksen number ($\ER\gg1$), which is probably more appropriate to the ODF method than the limit of small Ericksen number discussed in \cref{sec:app:LER}, we consider flow-aligning nematics \cite{Cousins2020}, again the most relevant choice for the materials used in the ODF method. In this case, at leading order the director is determined by the flow, and hence the anchoring conditions on both the free surface and the plates are not satisfied, the anchoring being broken by the flow effects.
Specifically, the torque on the director at the free surface and the plates due to anchoring forces is overcome by the torque due to flow effects, and the resulting director orientation is determined entirely by the flow.
This phenomenon is a flow-induced type of anchoring breaking that has been discussed in the context of channel flow by Cousins \etal \cite{Cousins2020}.

The solution in the limit of large Ericksen number for a flow-aligning nematic can be obtained from the equations given in the first column of \cref{table:HER}, with the solution to Poisson's equation for the pressure, which is given in \cref{ERinf:FA}, subject to the kinematic condition \cref{odf:bc} also providing the solutions for the velocities and twist director angle.
Similarly to in \cref{sec:app:LER}, we note that \cref{ERinf:FA,odf:bc} can be solved analytically using a conformal mapping (see \cite{Crowdy2001,Shelley1997} for more details); however, this approach again leads to integrals that must be evaluated numerically, and so again there is little advantage of pursuing this approach over a purely numerical approach.
We therefore again use COMSOL Multiphysics \cite{COMSOLcode} to numerically solve the Poisson equation for the pressure in \cref{ERinf:FA}.

\cref{fig:HER} shows the numerically calculated solutions for pressure $\tp$ and the director field $\n$ (top row), as well as the speed $|\vel|$ and streamlines (middle row), and the optical transmission $T$ (bottom row) at leading order in the limit of large Ericksen number ($\ER \gg 1$) for the parameters listed in \cref{table:ODF} at three different times: the first column, (a), (d) and (g), for $t= 0.02 \,$s when $d = 54\, \mu {\rm m}$ and $R' = 0.05\,$m$\,$s$^{-1}$; the second column, (b), (e) and (h), for $t=0.04 \,$s when $d = 34 \, \mu {\rm m}$ and $R' = 0.12\,$m$\,$s$^{-1}$; and the third column, (c), (f) and (i), for $t= 0.06 \,$s when $d= 14\, \mu {\rm m}$ and $R' = 0.49\,$m$\,$s$^{-1}$.
The pressure $\tp$ shown in \cref{fig:HER}(a)--(c) initially attains a local maximum at the centre of each droplet, and as $t$ increases, and hence as $d$ decreases, the maximum in the pressure moves towards the centre of the two coalescing droplets.
\cref{fig:HER}(d)--(f) shows that at $t=0.02$\,s the streamlines form two radial distributions with the streamline origins located at the maxima in the pressure, and as $t$ increases and the two maxima in the pressure approach the centre of the coalescing droplets, the streamlines approach a single radial distribution with the streamline origins located at the centre of the coalescing droplets.
As expected, the director field $\n$ shown in \cref{fig:HER}(a)--(c) is determined by the flow and hence aligns with the streamlines shown in \cref{fig:HER}(d)--(f).
Finally, \cref{fig:HER}(j)--(l) shows the optical transmission $T$ in which a $+1$-defect is positioned at the centre of each of the two coalescing droplets and a $-1$-defect is positioned at the centre of the two coalescing droplets \cite{GPBOOK}.
As $t$ increases, and hence as $d$ decreases, the defects move towards the centre of the two coalescing droplets and merge.
Again, we note that the optical transmission shown in \cref{fig:HER} (bottom row) provides a visualisation of the director field that can be readily compared to the optical transmission observed in experiments.

\section{Conclusions}
\label{sec:conc}

\begin{figure}[tp]
\begin{center}
\begin{tabular}{ccc}
\hspace{-0.3cm}
\includegraphics[angle=90,origin=c,height=0.3\linewidth]{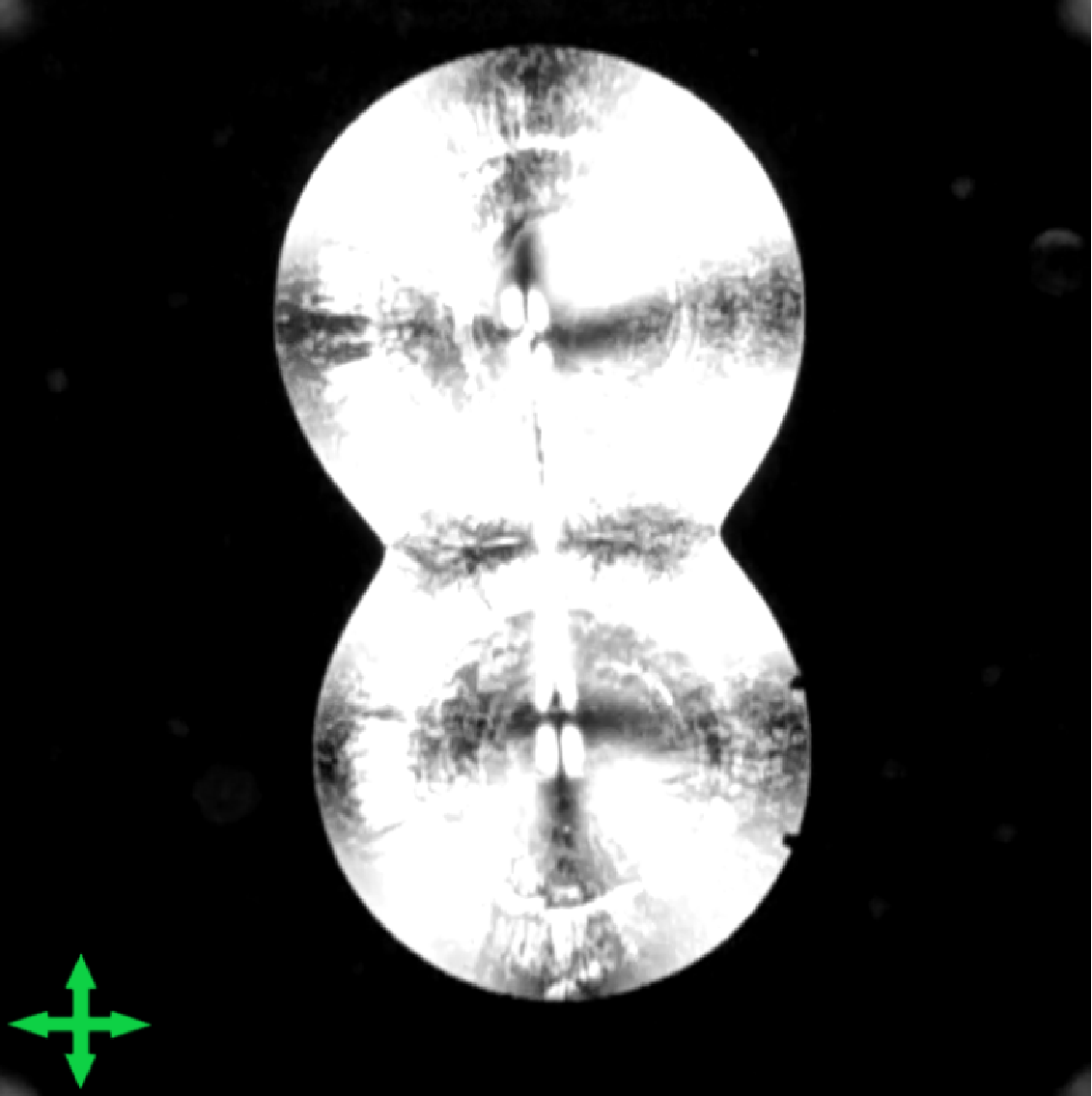} &
\hspace{-0.3cm}
\includegraphics[angle=90,origin=c,height=0.3\linewidth]{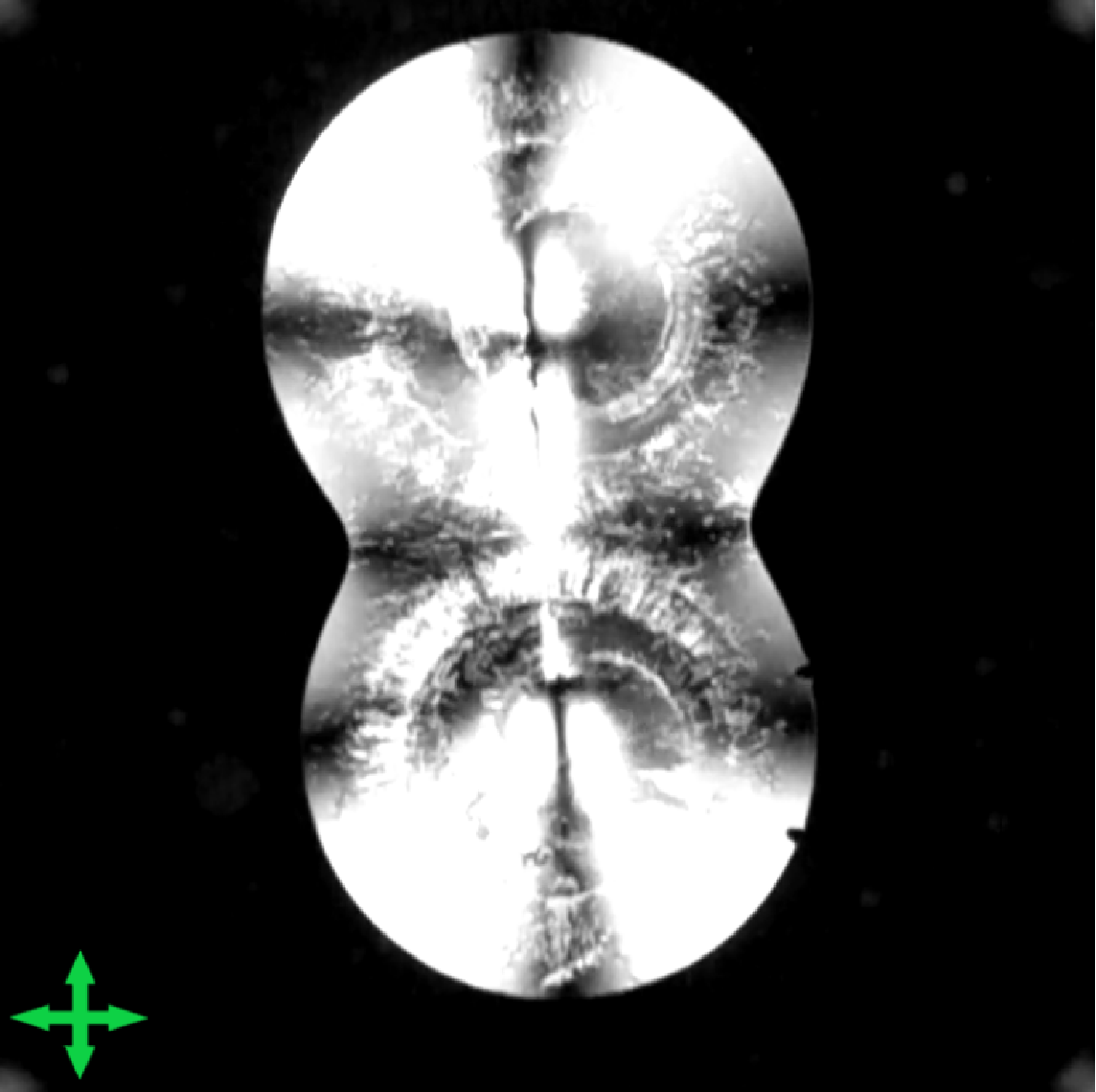} &
\hspace{-0.3cm}
\includegraphics[angle=90,origin=c,height=0.301\linewidth]{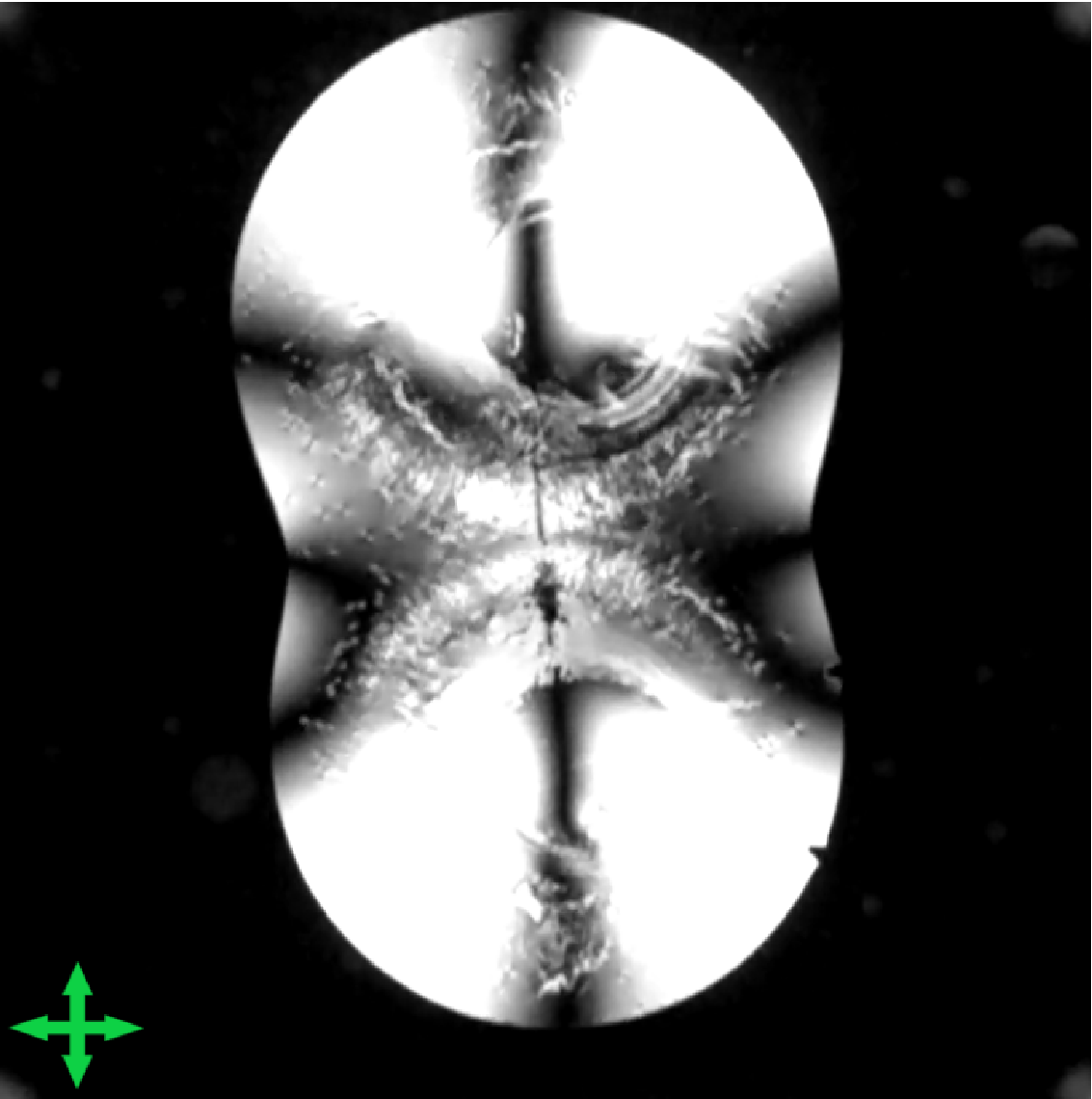}
\end{tabular}
\caption{
Experimental photographs of the optical transmission of light through a two-droplet ODF test setup between crossed polarisers by Merck KGaA using an unknown nematic material or nematic mixture (with polariser direction indicated by the green arrows). Left to right shows the increase of time.
Regions of white show complete optical transmission through the drops and regions of black show no optical transmission. [Photographs provided by Merck KGaA.]
}
\label{fig:Texp}
\end{center}
\end{figure}

In the present work, we considered the flow of a nematic in a standard Hele-Shaw cell that consists of two parallel plates, one of which may move in the direction perpendicular to the plates, separated by a narrow gap which is partially or wholly filled with the nematic.

In \cref{sec:prob,sec:EL,sec:BC,sec:BC:fs,sec:ND,sec:thinfilm}, we derived the thin-film Ericksen--Leslie equations that govern the flow and director within a nematic Hele-Shaw cell.
The thin-film Ericksen Leslie equations are given by the conservation of mass equation \cref{CMeq0}, the conservation of linear momentum equations \cref{LM1eq0,LM2eq0}, and the conservation of angular momentum equations \cref{AM1eq0,AM2eq0} subject to the no-slip and no-penetration conditions on the plates \cref{noslipHnd,noslip0nd} and we chose two general anchoring conditions described in \cref{sec:BC} that are relevant to a variety of situations.
These governing equations may, in principle, be generalised to include conservative body forces and other choices of boundary conditions on the plates and/or the free surface $\partial\Omega$.

In \cref{sec:ER0,sec:ERinf}, we solved the thin-film Ericksen--Leslie equations in the limits of small and large Ericksen numbers.
In the limit of small Ericksen number, the anchoring pattern on the plates determines the director field throughout the cell, and therefore the director field is fixed.
The fixed director field produces an anisotropic patterned viscosity. In particular, in the cases of unidirectional rubbed infinite anchoring with a constant pretilt and axisymmetric infinite anchoring with a constant pretilt, the flow is guided along the rubbing direction and guided by the axisymmetric anchoring pattern leading to a spiral flow, respectively.
Mi and Yang \cite{CapillaryMiYang} experimentally observed a reduction in the velocity of a nematic capillary flow when the flow was perpendicular to the rubbing direction compared to when the flow was parallel to the rubbing direction.
Analogous behaviour has also been observed in spreading nematic droplets on rubbed surfaces, both experimentally by Tortora and Lavrentovich \cite{Tortora2011} and in molecular dynamics simulations by Vanzo \etal \cite{Vanzo2016}, where elongation of the droplet free surface is guided by the rubbing direction, which leads to ellipsoidal nematic droplets known as tactoids.
In the limit of large Ericksen number, there are two cases, either the nematic is a flow-aligning nematic or the nematic is a non-flow-aligning nematic.
In both of these cases, the flow is identical to the flow of an isotropic fluid, and the behaviour of the director is determined by the flow.
A summary of the thin-film Ericksen–Leslie equations in the limits of small and large Ericksen numbers is given in \cref{table:LER,table:HER}, respectively.

Finally, in \cref{sec:app}, we applied the results of the limits of small and large Ericksen numbers to a simple model for the squeezing stage of the ODF method.
The optical transmission calculated in these limits, shown in \cref{fig:LER}(j)--(l) and \cref{fig:HER}(j)--(l), respectively, provides a clear visualisation of the director field which can be compared to the optical transmission observed in experiments.
The optical transmission of light measured through a two-droplet ODF test setup by Merck KGaA using an unknown nematic material or nematic mixture is shown in \cref{fig:Texp}.
Visual comparison of the experimental photographs and the results of the present theoretical model in the limit of large Ericksen number shown in \cref{fig:HER}(j)--(l) shows a striking resemblance, suggesting that the present theoretical model may provide a useful description of the ODF method.
We also note that it is clear from the experimental photographs shown in \cref{fig:Texp} that the droplets remain approximately cylindrical, justifying our modelling assumption that surface tension effects can be neglected.


We anticipate that many other nematic systems, including experiments on nematic viscous fingering and nematic microfluidics, can also be analysed using the thin-film Ericksen--Leslie equations derived in the present work, thus providing computationally cheaper models (which in some special cases allow for analytical solutions) than fully numerical alternatives, for studying flow in nematic Hele-Shaw cells.

\section*{Acknowledgements}

This work was supported by the U.K. Engineering and Physical Sciences Research Council (EPSRC), the University of Strathclyde, the University of Glasgow, and Merck KGaA via EPSRC research grants EP/P51066X/1 and EP/T012501/2.
The authors gratefully acknowledge Dr Brian R.\ Duffy of the University of Strathclyde, Dr Akhshay S.\ Bhadwal and Professor Carl V.\ Brown of Nottingham Trent University for discussions relating to this work, and Drs Leo Weegels of Merck KGaA and David Wilkes previously of Merck KGaA for discussions relating to LCD manufacturing.

\section*{Authors' ORCID Numbers}

Joseph R.\ L.\ Cousins 0000-0003-1723-5386,
Nigel J.\ Mottram 0000-0002-7265-0059, and
Stephen K.\ Wilson 0000-0001-7841-9643.

\appendix
\section{Nematic viscous dissipation}
\label{sec:appendixa}

The standard nematic viscous dissipation $\Disp$ \cite{ISBOOK}, given by \cref{Disp}, is obtained using the director and velocity in the form of \cref{director,velocity}, respectively, with the definitions of the co-rotational time flux of the director $\bm{N}$ and the rate of strain tensor $\bm{A}$ mentioned in the text. The components are summarised here:
\begin{align*}
    \big(\n \cdot \bm{A} \, \n\big)^2 &= \bigg[\cos^2\phi \cos^2\theta \,u_x+\cos^2\theta \cos\phi \sin\phi (u_y+v_x)+\cos^2\theta \sin^2\phi \,v_y+\cos\phi \cos\theta \sin\theta (u_z+w_x)\\
    &\quad+\cos\theta \sin\phi \sin\theta (v_z+w_y)+\sin^2\theta \,w_z\bigg]^2, \numberthis \\
    \bm{N} \cdot \bm{A}\,\n &= \dfrac{1}{4} \bigg[-2 \cos\phi \cos\theta u_x (2 \cos\theta \dot{\phi} \sin\phi+2 \cos\phi \dot{\theta} \sin\theta+\cos\theta \sin\phi u_y+\sin\theta u_z-\cos\theta \sin\phi v_x-\sin\theta w_x) \\
    &\quad -\cos\theta \sin\phi (u_y+v_x) (2 \cos\theta \dot{\phi} \sin\phi+2 \cos\phi \dot{\theta} \sin\theta+\cos\theta \sin\phi u_y+\sin\theta u_z-\cos\theta \sin\phi v_x-\sin\theta w_x)\\
    &\quad -\sin\theta (u_z+w_x) (2 \cos\theta \dot{\phi} \sin\phi+2 \cos\phi \dot{\theta} \sin\theta+\cos\theta \sin\phi u_y+\sin\theta u_z-\cos\theta \sin\phi v_x-\sin\theta w_x) \\
    &\quad+\cos^2\theta (\cos\phi (u_z+w_x)+\sin\phi (v_z+w_y)) (2 \dot{\theta}+\cos\phi u_z+\sin\phi v_z-\cos\phi w_x-\sin\phi w_y)\\
    &\quad -\sin\theta (v_z+w_y) (-2 \cos\phi \cos\theta \dot{\phi}+2 \dot{\theta} \sin\phi \sin\theta-\cos\phi \cos\theta u_y+\cos\phi \cos\theta v_x+\sin\theta v_z-\sin\theta w_y) \\
    &\quad +\cos\phi \cos\theta (u_y+v_x) (2 \cos\phi \cos\theta \dot{\phi}-2 \dot{\theta} \sin\phi \sin\theta+\cos\phi \cos\theta u_y-\cos\phi \cos\theta v_x-\sin\theta v_z+\sin\theta w_y) \\
    &\quad +2 \cos\theta \sin\phi v_y (2 \cos\phi \cos\theta \dot{\phi}-2 \dot{\theta} \sin\phi \sin\theta+\cos\phi \cos\theta u_y-\cos\phi \cos\theta v_x-\sin\theta v_z+\sin\theta w_y)\\
    &\quad+2 \cos\theta \sin\theta (2 \dot{\theta}+\cos\phi u_z+\sin\phi v_z-\cos\phi w_x-\sin\phi w_y) w_z\bigg], \numberthis \\
    \text{tr}(\bm{A}^2) &=\dfrac{1}{2} \bigg[2 u_x^2+(u_y+v_x)^2+2 v_y^2+(u_z+w_x)^2+(v_z+w_y)^2+2 w_z^2\bigg], \numberthis \\
    \big( \bm{A}\, \n \big)^2 &= \dfrac{1}{4} \bigg[4 \cos^2\phi \cos^2\theta \, u_x^2+4 \cos^2\theta \cos\phi \sin\phi \, u_x (u_y+v_x)+\cos^2\phi \cos^2\theta (u_y+v_x)^2 \\
    &\quad+\cos^2\theta \sin^2\phi (u_y+v_x)^2+4 \cos^2\theta \cos\phi \sin\phi (u_y+v_x) v_y+4 \cos^2\theta \sin^2\phi \, v_y^2\\
    &\quad +4 \cos\phi \cos\theta \sin\theta \, u_x (u_z+w_x)+2 \cos\theta \sin\phi \sin\theta (u_y+v_x) (u_z+w_x) \\
    &\quad+\cos^2\phi \cos^2\theta (u_z+w_x)^2+\sin^2\theta (u_z+w_x)^2+2 \cos\phi \cos\theta \sin\theta (u_y+v_x) (v_z+w_y)\\
    &\quad+4 \cos\theta \sin\phi \sin\theta v_y (v_z+w_y)+2 \cos^2\theta \cos\phi \sin\phi (u_z+w_x) (v_z+w_y)+\cos^2\theta \sin^2\phi (v_fz+w_y)^2 \\
    &\quad+\sin^2\theta (v_z+w_y)^2+4 \cos\phi \cos\theta \sin\theta (u_z+w_x) w_z\\
    &\quad+4 \cos\theta \sin\phi \sin\theta (v_z+w_y) w_z+4 \sin^2\theta w_z^2\bigg], \numberthis \\
    \bm{N}^2 &= \dfrac{1}{4} \bigg[(2 \cos\theta \,\dot{\phi} \,\sin\phi+2 \cos\phi \dot{\theta} \sin\theta+\cos\theta \sin\phi \,u_y+\sin\theta\, u_z-\cos\theta \sin\phi \,v_x-\sin\theta\, w_x)^2\\
    &\quad+\cos^2\theta (2 \dot{\theta}+\cos\phi\, u_z+\sin\phi \,v_z-\cos\phi\, w_x-\sin\phi \,w_y)^2\\
    &\quad+(2 \cos\phi \cos\theta \,\dot{\phi}-2 \dot{\theta} \sin\phi \sin\theta+\cos\phi \cos\theta\, u_y-\cos\phi \cos\theta \,v_x-\sin\theta \,v_z+\sin\theta \,w_y)^2\bigg], \numberthis
\end{align*}
where partial derivatives are denoted with subscripts, so that, for example, $u_x = \partial u/\partial x$.

\section{Typical parameter values}
\label{sec:appendixb}

The values of the timescales $\tau_1$, $\tau_2$ and $\tau_3$ and nondimensional numbers $\delta$, $\RE$ and $\ER$ in \cref{table:examples} are calculated from \cref{tau1,tau2,tau3,delta,Reynolds,Ericksen}, respectively, with the typical parameter values stated below.

Analysis of the ODF method was carried out by Cousins \etal \cite{Cousins2019} for a range of present and possible future lengthscales, depths and upper plate speeds given by $L = 50\,\mu$m--$5\,$mm, depths $D=50\,\mu$m--$0.5\,\mu$m, plate speeds $2 \, \mu$m\,s$^{-1}$--$1$mm\,s$^{-1}$ (which yields a velocity scale in the range $U = 0.0002\,$--$0.1$ m\,s$^{-1}$), respectively.
Typical values for the nematic isotropic viscosity $\mu = 0.01\,$Pas, one-constant elastic constant $K = 10\,$pN, and density $\rho = 1000\,$kg\,m$^{-3}$ are also used.

Capillary-filling experiments were carried out by Mi and Yang \cite{CapillaryMiYang} for the nematic 5CB in rectangular channels of length $L = 0.1\,$m, depth $D=10\,\mu$m, with front propagation speed of $U = 0.01\,$cm$\,$s$^{-1}$ for the nematic 5CB \cite{CapillaryMiYang}.

Air--nematic viscous fingering experiments were carried out by Sonin and Bartolino \cite{Sonin1993} for the nematic 5CB by lifting the upper plate of a Hele-Shaw cell of length $L=0.1\,$mm, depth $D=3.5\,\mu$m, upward plate speed $280\,\mu$m\,s$^{-1}$ (which yields a velocity scale of $U = 8\,$mm\,s$^{-1}$).

Microfluidic experiments in channels containing a micropillar were carried out by Sengupta \etal \cite{Sengupta2013d} for the nematic 5CB in channels of length $L = 20\,$mm, with a range of depths $D = 8$--$50\mu$mm, and flow velocities $U=200$--$670\,\mu$m\,s$^{-1}$.

\bibliography{bib}

\end{document}